\documentclass[prd,twocolumn,superscriptaddress,altaffilletter,nofootinbib]{revtex4}

\usepackage{cancel}
\usepackage{graphicx}
\usepackage{amsmath}
\usepackage{amssymb}
\usepackage{graphicx,epsfig}
\newtheorem{theorem}{Theorem}
\newtheorem{corollary}{Corollary}[theorem]

\newcommand{\be}{\begin{equation}}
\newcommand{\ee}{\end{equation}}
\newcommand{\bea}{\begin{eqnarray}}
\newcommand{\eea}{\end{eqnarray}}
\newcommand{\der}{\partial}
\newcommand{\vphi}{\varphi}
\newcommand{\bet}{\begin{theorem}}
\newcommand{\eet}{\end{theorem}}
\newcommand{\bec}{\begin{corollary}}
\newcommand{\eec}{\end{corollary}}

\begin{document}



\title{Gauge invariant approach to nonmetricity theories and the second clock effect}



\author{Israel Quiros}\email{iquiros@fisica.ugto.mx}\affiliation{Dpto. Ingenier\'ia Civil, Divisi\'on de Ingenier\'ia, Universidad de Guanajuato, Gto., M\'exico.}



\begin{abstract}
In this paper we discuss on recent attempts aimed at demonstrating that, contrary to well-known results, the second clock effect (SCE) does not take place in generalized Weyl spaces -- spaces with arbitrary nonmetricity -- denoted here as $W_4$ spaces. These attempts include Weyl gauge theories of gravity, as well as the symmetric teleparallel theories (STTs). Our approach to this issue is based on the adoption of Weyl gauge symmetry (WGS) which is a manifest symmetry of the basic laws of Weyl geometry. We shall consistently adapt mathematical and geometrical quantities and concepts so that the resulting geometrical framework be gauge invariant. This issue is of special relevance for the fate of nonmetricity theories, including a class of the STTs which is being intensively applied in the cosmological framework. As we shall show, if realize that WGS is a manifest symmetry of generalized Weyl spaces $W_4$, and identify physical vectors and tensors with corresponding hypothetical vectors and tensors living in $W_4$, neither the Weyl gauge theories nor the nonmetricity theories are free of the SCE, unless Weyl integrable geometry (WIG) spaces are considered.
\end{abstract}



\maketitle


\section{Introduction} 

In general relativity (GR) when two identical clocks, initially synchronized, are parallel transported along different paths, a certain loss of synchronization arises that is called as the ``first clock effect.'' In Weyl spacetimes, where the following ``vectorial'' nonmetricity condition is satisfied \cite{delhom-2019}:

\bea \nabla_\alpha g_{\mu\nu}=-Q_\alpha g_{\mu\nu},\label{weyl-nm}\eea with $Q_\alpha$ -- the Weyl gauge vector, an additional effect arises: the two clocks not only have lost their initial synchronization, but, they go at different rates. It is known as the second clock effect \cite{adler-book, perlick-1991, tucker, 2clock-1, 2clock-2, 2clock-4, 2clock-5, many-weyl-book, 2clock-tomi, tomi-replay, 2c-delhom, hobson, hobson-replay}. The SCE causes a serious observational issue: an unobserved broadening of spectral lines. This issue was enough to reject the original Weyl's gauge invariant gravitational theory and its related geometrical framework \cite{weyl-1917}. 

Despite of this well-known result, in recent papers the occurrence of the second clock effect has been challenged \cite{2clock-tomi, tomi-replay, hobson, hobson-replay}. It has been demonstrated in \cite{2clock-tomi} a lemma (lemma 2) which basically states that, in spaces with generalized nonmetricity $Q_{\alpha\mu\nu}$, where the teleparallel condition (vanishing generalized curvature) is satisfied, the SCE does not take place. In that reference, however, the authors did not consider the WGS which is a manifest symmetry of generalized Weyl spaces and, besides, as we shall show in section \ref{sect-challenge} of the present paper, they considered an statement of the parallel transport law which is not valid for tangent vectors with conformal weight $-1$. In consequence, the main equation in section 5 of \cite{2clock-tomi} -- equation (15) -- is not valid when any of the vectors in the inner product has weight $-1$. In addition, the demonstration starts with a path integral along a closed timelike curve (CTC) -- see equation (12) of the above reference -- which may bring into consideration causality issues, as it is discussed in section \ref{sect-ctc} of the present paper.

In reference \cite{hobson} the authors demonstrated that if one takes into account the Weyl gauge symmetry, and one requires, besides, the presence of massive matter fields to represent atoms, observers and clocks, then Weyl gauge theories do not predict a SCE. This result is interesting, besides, because it was obtained in standard Weyl geometry spaces, where it is supposed that the occurrence of the SCE was demonstrated long ago. However, as we shall show below in section \ref{sect-p-transp}, the assumptions made in \cite{hobson, hobson-replay} amount to deny relating physical vectors and tensors with corresponding (hypothetical) vectors and tensors living in generalized Weyl space $W_4$. In other words, these assumptions amount to giving up the description of physical phenomena in $W_4$ space.

In this paper we shall show that WGS is not only a manifest symmetry of standard Weyl space $\tilde W_4$, but it is also a manifest symmetry of generalized Weyl spaces $W_4$, thus confirming previous results \cite{delhom-2019}. This result is possible after the development in section \ref{sect-p-transp}, of the gauge invariant theory of parallel transport, including the related concepts of gauge derivative along given path, etc. Here we adopt a parallel transport law that differs from the one undertaken in \cite{my-arxiv-paper}. Yet, similar results regarding to the occurrence of the second clock effect, are obtained in the present paper. There are other issues of current interest that we shall investigate in this paper as well. For instance: i) is the generalization of nonmetricity $Q_{\alpha\mu\nu}$, phenomenologically viable? and ii) do the matter fields interact with nonmetricity? The answers to the above questions carry important consequences for the nonmetricity theories, including the symmetric teleparallel theories of gravity. 

Our main goal will be to demonstrate that, if properly take into account the WGS and, besides, identify the physical vectors and tensors with the corresponding hypothetical vectors and tensors in $W_4$, the SCE must necessarily take place in generalized Weyl spaces. These include as subclasses the standard Weyl geometry spaces and those spacetimes where the teleparallel condition is satisfied. Our demonstration contradicts recent results in \cite{2clock-tomi} and in \cite{hobson, hobson-replay} and confirms well-known (long standing) results. Although this paper is not intended as a comment on the mentioned references, here we shall show why the corresponding results should be taken with caution. 

All of the results discussed here are based on the assumption of a parallel transport law that differs from the one assumed in \cite{my-arxiv-paper} and also, in a specific consistency hypothesis that allows identification of physical vectors and tensors with related hypothetical vectors and tensors living in $W_4$ space. Through the paper, for simplicity, we consider only Riemann-Christoffel and nonmetricity contributions to the curvature and affinity of space. The torsion contribution is omitted not only for simplicity of the analysis but, also, because consideration of spaces with torsion is beyond the scope of our planned discussion. 

We have organized this paper in the following way. Sections \ref{sect-basic} and \ref{sect-wgs} are introductory. In section \ref{sect-basic} the basic notions of generalized Weyl space -- spaces with Riemann-Christoffel curvature and arbitrary nonmetricity -- as well as our conventions, are given. Then, in section \ref{sect-wgs}, the rudiments of generalized Weyl gauge symmetry are exposed. Sections \ref{sect-p-transp} and \ref{sect-geod} are dedicated mostly to expose the mathematical developments behind a consistent statement of gauge invariant parallel transport which is alternative to the one assumed in \cite{my-arxiv-paper}. The parallel transport law is required to understand the second clock effect and related subjects covered in this paper. In section \ref{sect-p-transp}, in particular, the master equations that serve as the mathematical basis for the SCE, are derived. In the second part of the paper, which is composed of sections \ref{sect-sce}, \ref{sect-challenge}, \ref{sect-fermions} and \ref{sect-ctc}, we deal with the geometrical and phenomenological consequences of our gauge invariant approach to generalized Weyl spaces. It is in these sections where we demonstrate the inevitability of the SCE in $W_4$ as a consequence of: (i) the assumed parallel transport law and (ii) the consistency requirement that allows to identify physical vectors and tensors with the related hypothetical vectors and tensors in generalized Weyl space $W_4$. Besides, in these sections we explore the answers to the questions stated in the former paragraphs as, for instance: Do the spinor matter fields and spinning test bodies interact with nonmetricity? which is investigated in section \ref{sect-fermions}. In section \ref{sect-discu} we discuss on the results obtained, while concluding remarks are given in section \ref{sect-conclu}. For completeness an appendix section has been included. In appendix \ref{app-a} we discuss on what happens if consider gauge symmetry in a situation like the one discussed in lemma 2 of \cite{2clock-tomi}, while in appendix \ref{app-b} we have included a brief reply to a comment appeared in reference \cite{tomi-replay}.

In this paper, unless otherwise stated, we use the units $\hbar=c=1$ and the following signature of the metric is chosen: $(-+++)$.



\section{Background and conventions}\label{sect-basic} 

The generalized Weyl geometry \cite{delhom-2019}, denoted here by $W_4$, is defined as the class of four-dimensional (torsionless) manifolds ${\cal M}_4$ that are paracompact, Hausdorff, connected $C^\infty$, endowed with a locally Lorentzian metric $g$ that obeys the following nonmetricity condition:

\bea \nabla_\mu g_{\mu\nu}=-Q_{\alpha\mu\nu},\label{gen-nm}\eea where $Q_{\alpha\mu\nu}$ is the nonmetricity tensor and the covariant derivative $\nabla_\mu$ is defined with respect to the generalized torsion-free affine connection of the manifold: 

\bea \Gamma^\alpha_{\;\;\mu\nu}=\{^\alpha_{\mu\nu}\}+L^\alpha_{\;\;\mu\nu}\;\stackrel{\text{symb.}}{\longrightarrow}\;\Gamma=\{\}+L,\label{gen-aff-c}\eea where 

\bea \{^\alpha_{\mu\nu}\}:=\frac{1}{2}g^{\alpha\lambda}\left(\der_\nu g_{\mu\lambda}+\der_\mu g_{\nu\lambda}-\der_\lambda g_{\mu\nu}\right),\label{lc-aff-c}\eea is the Levi-Civita (LC) connection, while 

\bea L^\alpha_{\;\;\mu\nu}:=\frac{1}{2}\left(Q^{\;\;\alpha}_{\mu\;\;\nu}+Q^{\;\;\alpha}_{\nu\;\;\mu}-Q^\alpha_{\;\;\mu\nu}\right),\label{disf-t}\eea is the disformation tensor. The nonmetricity tensor $Q_{\alpha\mu\nu}$ is symmetric in the second and third indices. It measures how much the length of given vector varies during parallel transport \cite{beltran-univ-2019}.

Standard Weyl geometry, denoted here by $\tilde W_4$, is a subclass of $W_4$ which is defined by the choice of vectorial nonmetricity $Q_{\alpha\mu\nu}=Q_\alpha g_{\mu\nu}$, and so gives rise to \eqref{weyl-nm}, where the operator $\nabla_\alpha$ denotes covariant derivative defined with respect to the connection \eqref{gen-aff-c}, but with the disformation tensor given by:

\bea L^\alpha_{\;\;\mu\nu}:=\frac{1}{2}\left(Q_\mu\delta^\alpha_\nu+Q_\nu\delta^\alpha_\mu-Q^\alpha g_{\mu\nu}\right),\label{disf-t'}\eea instead of \eqref{disf-t}.

In this paper we call as ``generalized curvature tensor'' of $W_4$ spacetime, the curvature of the connection, symbolically ${\bf R}(\Gamma)$, whose coordinate components are given by:

\bea &&R^\alpha_{\;\;\sigma\mu\nu}:=\der_\mu\Gamma^\alpha_{\;\;\nu\sigma}-\der_\nu\Gamma^\alpha_{\;\;\mu\sigma}\nonumber\\
&&\;\;\;\;\;\;\;\;\;\;\;\;\;\;\;+\Gamma^\alpha_{\;\;\mu\lambda}\Gamma^\lambda_{\;\;\nu\sigma}-\Gamma^\alpha_{\;\;\nu\lambda}\Gamma^\lambda_{\;\;\mu\sigma},\label{gen-curv-t}\eea or, if take into account the decomposition \eqref{gen-aff-c}:

\bea &&R^\alpha_{\;\;\sigma\mu\nu}=\hat R^\alpha_{\;\;\sigma\mu\nu}+\hat\nabla_\mu L^\alpha_{\;\;\nu\sigma}-\hat\nabla_\nu L^\alpha_{\;\;\mu\sigma}\nonumber\\
&&\;\;\;\;\;\;\;\;\;\;\;\;\;\;+L^\alpha_{\;\;\mu\lambda}L^\lambda_{\;\;\nu\sigma}-L^\alpha_{\;\;\nu\lambda}L^\lambda_{\;\;\mu\sigma},\label{gen-curv-t-1}\eea where $\hat R^\alpha_{\;\;\sigma\mu\nu}$ is the Riemann-Christoffel or LC curvature tensor,

\bea &&\hat R^\alpha_{\;\;\sigma\mu\nu}:=\der_\mu\{^\alpha_{\nu\sigma}\}-\der_\nu\{^\alpha_{\mu\sigma}\}\nonumber\\
&&\;\;\;\;\;\;\;\;\;\;\;\;\;\;\;+\{^\alpha_{\mu\lambda}\}\{^\lambda_{\nu\sigma}\}-\{^\alpha_{\nu\lambda}\}\{^\lambda_{\mu\sigma}\},\label{lc-curv-t}\eea and $\hat\nabla_\alpha$ is the LC covariant derivative. Besides, the LC Ricci tensor $\hat R_{\mu\nu}=\hat R^\lambda_{\;\;\mu\lambda\nu}$ and LC curvature scalar read:

\bea &&\hat R_{\mu\nu}=\der_\lambda\{^\lambda_{\nu\mu}\}-\der_\nu\{^\lambda_{\lambda\mu}\}+\{^\lambda_{\lambda\kappa}\}\{^\kappa_{\nu\mu}\}-\{^\lambda_{\nu\kappa}\}\{^\kappa_{\lambda\mu}\},\nonumber\\
&&\hat R=g^{\mu\nu}\hat R_{\mu\nu},\label{lc-curv-sc}\eea respectively. We call $R^\alpha_{\;\;\sigma\mu\nu}$ as generalized curvature tensor because it is contributed both by LC curvature $\hat R^\alpha_{\;\;\sigma\mu\nu}$, and by nonmetricity through disformation $L^\alpha_{\;\;\mu\nu}$. We have that,

\bea &&R_{\mu\nu}=\hat R_{\mu\nu}+\hat\nabla_\lambda L^\lambda_{\;\;\mu\nu}-\hat\nabla_\nu L^\lambda_{\;\;\lambda\mu}\nonumber\\
&&\;\;\;\;\;\;\;\;\;\;\;+L^\lambda_{\;\;\lambda\kappa}L^\kappa_{\;\;\mu\nu}-L^\lambda_{\;\;\nu\kappa}L^\kappa_{\;\;\lambda\mu},\label{gen-ricci-t}\\
&&R=\hat R+{\cal Q}+\der{\cal Q},\label{gen-curv-sc}\eea where the nonmetricity scalar ${\cal Q}$ and the boundary term $\der{\cal Q}$, are defined as it follows:

\bea &&{\cal Q}:=L^\tau_{\;\;\tau\lambda}L^{\lambda\kappa}_{\;\;\;\;\kappa}-L^{\tau\kappa\lambda}L_{\lambda\tau\kappa},\nonumber\\
&&\der{\cal Q}:=\hat\nabla_\lambda\left(L^{\lambda\kappa}_{\;\;\;\;\kappa}-L^{\;\;\kappa\lambda}_\kappa\right).\label{q-inv}\eea 

If take into consideration the teleparallel condition (see equation \eqref{telep-cond} below), equation \eqref{gen-curv-sc} is the mathematical basis for the claimed equivalence between GR and STT.


\subsection{Properties and identities of the curvature in $W_4$}


For the (torsionless) connection $\nabla$ of $W_4$ space it is verified the second Bianchi identity:

\bea \nabla_\mu R^\kappa_{\;\;\lambda\nu\sigma}+\nabla_\nu R^\kappa_{\;\;\lambda\sigma\mu}+\nabla_\sigma R^\kappa_{\;\;\lambda\mu\nu}=0.\label{2-bianchi}\eea From this identity, taking into account that $$\nabla_\alpha g_{\mu\nu}=-Q_{\alpha\mu\nu},\;\nabla_\alpha g^{\mu\nu}=Q_\alpha^{\;\;\mu\nu},\;Q^\mu_{\;\;\mu\nu}=Q_\nu,$$ etc., we obtain the following equation:

\bea &&\nabla^\nu G_{\nu\alpha}=\frac{1}{2}\left(Q_\alpha g^{\mu\nu}-Q_\alpha^{\;\;\mu\nu}\right)R_{\mu\nu}\nonumber\\
&&\;\;\;\;\;\;\;\;\;\;\;\;\;\;\;+\frac{1}{2}\left(Q_\lambda^{\;\;\mu\nu}-Q_\lambda g^{\mu\nu}\right)R^\lambda_{\;\;\mu\alpha\nu},\label{2-bianchi-einstein-t}\eea where $G_{\mu\nu}\equiv R_{\mu\nu}-g_{\mu\nu}R/2$ is the Einstein's tensor of $W_4$ space. This equation amounts to a generalization of the Bianchi identity of the Einstein's tensor. In standard Weyl geometry space, since $$Q_\alpha g^{\mu\nu}-Q_\alpha^{\;\;\mu\nu}=\left(Q_\alpha-Q_\alpha\right)g^{\mu\nu}=0,$$ the well-known Bianchi identity of the generalized Einstein's tensor: $\nabla^\nu G_{\nu\alpha}=0$, is recovered.

The symmetries of the generalized curvature tensor in $W_4$ space differ from those in Riemann space $V_4$. For instance:

\bea &&R^\alpha_{\;\;\sigma\mu\nu}=-R^\alpha_{\;\;\sigma\nu\mu},\label{symm-1-gcurv}\\
&&R_{\alpha\sigma\mu\nu}=-R_{\sigma\alpha\mu\nu}+\nabla_\mu Q_{\nu\alpha\sigma}-\nabla_\nu Q_{\mu\alpha\sigma}.\label{symm-2-gcurv}\eea The last equation, known as the third Biancchi identity, in compact form can be written in the following way \cite{2clock-tomi}:

\bea \nabla_{[\mu}Q_{\nu]\alpha\sigma}=R_{(\alpha\sigma)\mu\nu}.\label{3-bianchi-id}\eea 

In standard Weyl geometry where $Q_{\alpha\mu\nu}=Q_\alpha g_{\mu\nu}$, so that $Q^\mu_{\;\;\mu\nu}=Q_\nu$ and $Q^{\;\;\mu}_{\nu\;\;\mu}=4Q_\nu$), 

\bea &&\nabla_\mu Q_{\nu\alpha\sigma}-\nabla_\nu Q_{\mu\alpha\sigma}=\left(\nabla_\mu Q_\nu-\nabla_\nu Q_\mu\right)g_{\alpha\sigma}\nonumber\\
&&\;\;\;\;\;\;\;\;\;\;\;\;\;\;\;\;\;\;\;\;\;\;\;\;\;\;\;\;\;\;\;\;=\left(\der_\mu Q_\nu-\der_\nu Q_\mu\right)g_{\alpha\sigma},\nonumber\eea the property \eqref{symm-2-gcurv} can be written in the following way:

\bea \der_{[\mu}Q_{\nu]}g_{\alpha\sigma}=R_{(\alpha\sigma)\mu\nu}.\label{3-bianchi'}\eea Besides, in the particular case when $Q_\alpha=\der_\alpha\phi$ ($\phi$ is a scalar function), since $\der_\mu Q_\nu-\der_\nu Q_\mu=0$, the generalized curvature tensor possesses the same symmetries of the tensor indices as the Riemann-Christoffel curvature tensor. This special case of Weyl geometry is known in the bibliography as Weyl integrable geometry or WIG for short.


\subsection{Symmetric teleparallel spacetimes}\label{sect-stt}

We shall call as teleparallel Weyl space $Z_4$, a paracompact, Hausdorff, connected $C^\infty$ four-dimendional manifold ${\cal M}_4$, endowed with a locally Lorentzian metric $g$ and the generalized (torsionless) connection $\Gamma$ decomposed as in \eqref{gen-aff-c} and satisfying \cite{beltran-univ-2019, hohmann-2021}:

\bea \Gamma^\alpha_{\;\;\mu\nu}=\{^\alpha_{\mu\nu}\}+L^\alpha_{\;\;\mu\nu}=(\Lambda^{-1})^\alpha_{\;\;\lambda}\der_\mu\Lambda^\lambda_{\;\;\nu},\label{gen-aff-c-z4}\eea where $\Lambda^\mu_{\;\;\nu}$ is an element of the general linear group $GL(4,\mathbb{R})$ \cite{beltran-univ-2019} and $(\Lambda^{-1})^\alpha_{\;\;\lambda}$ is its inverse, so that: $(\Lambda^{-1})^\mu_{\;\;\lambda}\Lambda^\lambda_{\;\;\nu}=\delta^\mu_\nu.$ The connection $\Gamma$ is purely inertial. In the absence of torsion this form of the connection leads to the additional constraint $\der_{[\mu}\Lambda^\alpha_{\;\;\nu]}=0$. Hence, the general element of $GL(4,\mathbb{R})$ determining the connection can be parametrized by a set of functions $\chi^\mu$ so that \cite{beltran-univ-2019}:

\bea \Gamma^\alpha_{\;\;\mu\nu}=\frac{\der x^\alpha}{\der\chi^\lambda}\der_\mu\der_\nu\chi^\lambda.\label{univ-2019}\eea

If substitute the connection \eqref{univ-2019} into the definition \eqref{gen-curv-t} of the generalized curvature tensor, one obtains another expression for the teleparallel condition:

\bea {\bf R}(\Gamma)=0\;\Leftrightarrow\;R^\alpha_{\;\;\sigma\mu\nu}=0.\label{telep-cond}\eea Hence, teleparallel Weyl space $Z_4$ is the same as flat $W_4$ space: 

\bea W_4\stackrel{{\bf R}(\Gamma)=0}{\longrightarrow}Z_4.\label{w4-z4}\eea

The constraint \eqref{univ-2019} further restricts the fields $\chi^\mu$, the metric and the nonmetricity to satisfy (see \cite{beltran-univ-2019} for a similar relationship in the absence of nonmetricity): 

\bea \der_\alpha g_{\mu\nu}+Q_{\alpha\mu\nu}=2\frac{\der x^\lambda}{\der\chi^\sigma}\der_\alpha\der_{(\mu}\chi^\sigma g_{\nu)\lambda}.\label{telep-rel}\eea Under a specific choice of the field $\chi^\mu$, the above equation determines fixed relationships between the derivative of the metric and nonmetricity tensor. 

In what follows we shall explore the consequences of Weyl gauge symmetry for generalized Weyl spaces, which include the teleparallel spaces as a particular class.



\section{Weyl gauge symmetry}\label{sect-wgs}
 

WGS or invariance under local changes of scale \cite{hobson}, is a manifest symmetry of generalized Weyl geometry spaces $W_4$. In reference \cite{delhom-2019} an alternative definition of $W_4$ space is given, where the WGS is made evident (definition 3 of the mentioned reference): ``A conformally generalized Weyl structure is a differentiable manifold $\cal M$ endowed with a unique torsion-free affine connection and a conformally related equivalence class of metric tensors.'' Here we shall give a very compact exposition of this subject.

The geometric laws that define $W_4$ (also $\tilde W_4$ and $Z_4$), among which is the nonmetricity condition \eqref{gen-nm}, are invariant under generalized (local) Weyl rescalings or, also, Weyl gauge transformations. These amount to simultaneous conformal transformations of the metric\footnote{Here we assume that the conformal transformation of the metric in the Weyl gauge transformations, does not represent a diffeomorphism or, properly, a conformal isometry. Moreover, the spacetime coincidences or events, as well as the spacetime coordinates that label the points in spacetime, are not modified or altered by the conformal transformations in any way.\label{foot-1}} and gauge transformations of nonmetricity \cite{delhom-2019, saridakis-prd-2020, my-arxiv-paper}:

\bea &&g_{\mu\nu}\rightarrow\Omega^2g_{\mu\nu},\;g^{\mu\nu}\rightarrow\Omega^{-2}g^{\mu\nu},\nonumber\\
&&Q^\alpha_{\;\;\mu\nu}\rightarrow Q^\alpha_{\;\;\mu\nu}-2\der^\alpha\ln\Omega\,g_{\mu\nu},\nonumber\\
&&Q^{\;\;\alpha}_{\mu\;\;\nu}\rightarrow Q^{\;\;\alpha}_{\mu\;\;\nu}-2\der_\mu\ln\Omega\,\delta^\alpha_\nu,\nonumber\\
&&Q_{\alpha\mu\nu}\rightarrow\Omega^2\left(Q_{\alpha\mu\nu}-2\der_\alpha\ln\Omega\,g_{\mu\nu}\right),\nonumber\\
&&Q_\alpha\rightarrow Q_\alpha-2\der_\alpha\ln\Omega,\label{gauge-t}\eea respectively, where the positive smooth function $\Omega^2\equiv\Omega^2(x)$ is the conformal factor and we have identified 

\bea Q_\alpha\equiv Q^\lambda_{\;\;\lambda\alpha}=Q^\lambda_{\;\;\alpha\lambda}.\label{q-vec}\eea 

In what follows we shall call the transformations \eqref{gauge-t} either as Weyl gauge transformations or, simply, as gauge transformations. Under \eqref{gauge-t}:

\bea &&\{^\alpha_{\mu\nu}\}\rightarrow\{^\alpha_{\mu\nu}\}+\left(\delta^\alpha_\mu\der_\nu+\delta^\alpha_\nu\der_\mu-g_{\mu\nu}\der^\alpha\right)\ln\Omega,\nonumber\\
&&L^\alpha_{\;\;\mu\nu}\rightarrow L^\alpha_{\;\;\mu\nu}-\left(\delta^\alpha_\mu\der_\nu+\delta^\alpha_\nu\der_\mu-g_{\mu\nu}\der^\alpha\right)\ln\Omega,\label{connect-t}\eea so that the generalized affine connection \eqref{gen-aff-c} is unchanged by the Weyl rescalings:

\bea \Gamma^\alpha_{\;\;\mu\nu}\rightarrow\Gamma^\alpha_{\;\;\mu\nu}.\label{aff-c-t}\eea This means that the generalized curvature tensor $R^\alpha_{\;\;\sigma\mu\nu}$ in \eqref{gen-curv-t} and the generalized Ricci tensor, $R_{\mu\nu}\equiv R^\lambda_{\;\;\mu\lambda\nu}$, are unchanged as well, while the generalized curvature scalar transforms as it follows:

\bea R^\alpha_{\;\;\mu\sigma\nu}\rightarrow R^\alpha_{\;\;\mu\sigma\nu},\;R_{\mu\nu}\rightarrow R_{\mu\nu}\Rightarrow\;R\rightarrow\Omega^{-2}R.\label{curv-t-conf-t}\eea It can be straightforwardly demonstrated, as well, that the third Bianchi identity \eqref{3-bianchi-id} is a gauge invariant expression.

Gauge symmetry in $\tilde W_4$ spaces, which are characterized by vectorial non-metricity: $Q_{\alpha\mu\nu}=Q_\alpha g_{\mu\nu}$, amounts to invariance (or covariance) under:

\bea g_{\mu\nu}\rightarrow\Omega^2g_{\mu\nu},\;\;Q_\mu\rightarrow Q_\mu-2\der_\mu\ln\Omega.\label{conf-t}\eea


\subsection{WGS and STTs}

Due to the claimed equivalence between GR and symmetric teleparallel equivalent of general relativity (STEGR) \cite{beltran-univ-2019}, which is the simplest of the STTs, it may be argued that WGS must not be a symmetry of STTs. In this regard let us point out that, on the one hand, the mentioned equivalence is based on equation:

\bea &&\hat R^\alpha_{\;\;\sigma\mu\nu}=\hat\nabla_\nu L^\alpha_{\;\;\mu\sigma}-\hat\nabla_\mu L^\alpha_{\;\;\nu\sigma}\nonumber\\
&&\;\;\;\;\;\;\;\;\;\;\;\;\;\;+L^\alpha_{\;\;\nu\lambda}L^\lambda_{\;\;\mu\sigma}-L^\alpha_{\;\;\mu\lambda}L^\lambda_{\;\;\nu\sigma},\label{gen-curv-t-0}\eea which follows from \eqref{gen-curv-t-1} under the teleparallel requirement $R^\alpha_{\;\;\sigma\mu\nu}=0$. From \eqref{gen-curv-t-0} it follows that, 

\bea \hat R=-{\cal Q}-\der{\cal Q}.\label{equiv}\eea 

On the other hand the teleparallel condition \eqref{gen-aff-c-z4}, or in the form $R^\alpha_{\;\;\sigma\mu\nu}=0$, is invariant under \eqref{gauge-t}. Moreover, equations \eqref{gen-curv-t-0} and \eqref{equiv}, on which the mentioned equivalence between GR and pure nonmetricity theories of gravity is based, are Weyl gauge invariant as well. We should not forget that the dynamical equivalence between GR and STEGR takes place in flat $W_4$ space \eqref{w4-z4}, i. e., in teleparallel $Z_4$ space, which is manifest gauge symmetric. Hence, WGS being a manifest symmetry of $Z_4$ spaces, should be shared at least as an implicit symmetry, by STTs and by GR.\footnote{General relativity may be understood as a particular gauge of a gauge invariant theory of gravity \cite{my-arxiv-14}.} Take, for instance, the coincident gauge of teleparallel $Z_4$ space, where the connection vanishes \cite{beltran-univ-2019, coincident}: 

\bea \Gamma^\alpha_{\;\;\mu\nu}=0\;\Rightarrow\;\{^\alpha_{\mu\nu}\}=-L^\alpha_{\;\;\mu\nu},\label{coincident-cond}\eea i. e., we have that,

\bea \der_\alpha g_{\mu\nu}=-Q_{\alpha\mu\nu}.\label{coincident-cond-1}\eea 

Equations \eqref{coincident-cond} and \eqref{coincident-cond-1} are obviously invariant under \eqref{gauge-t}. Hence, WGS should play an important role in the development of the STTs as well. In this regard, in \cite{saridakis-prd-2020}, a family of conformal -- thus gauge invariant -- theories with second-order field equations and having the metric tensor as the fundamental variable, was formulated within the symmetric teleparallel framework. Our discussion in the present paper must be of importance, at least, for this family of theories. The $f({\cal Q})$ theories, on the other hand, are not gauge invariant in general, so that, in these cases, the gravitational Lagrangian does not share gauge invariance of $Z_4$ space and one may dispense with gauge symmetry.



\section{Gauge symmetry and parallel transport}\label{sect-p-transp}


Parallel transport consistent with gauge symmetry is required to define gauge covariant differentiation of vectors and tensors in generalized Weyl spaces. Below we shall develop the theory of gauge invariant parallel transport, which is cornerstone to discuss on the SCE. Although the present theory contains new elements not previously considered, to a great extent it is based in the work of references \cite{dirac-1973, maeder-1978}. Here we assume a statement of the parallel transport law that slightly differs from the one assumed in the latter references and in \cite{my-arxiv-paper}.



\subsection{Gauge derivative operators}\label{subsect-g-diff} 

In order to make the gauge symmetry compatible with well-known derivation rules and with the inclusion of fields into $W_4$, it is necessary to introduce the Weyl gauge derivative operators in a way that is equivalent to the one appearing in \cite{dirac-1973, utiyama-1973, maeder-1978}. Let ${\bf T}$ be a tensor in $W_4$, with coordinate components $T^{\alpha_1\alpha_2\cdots\alpha_i}_{\beta_1\beta_2\cdots\beta_j}$ and with conformal weight $w({\bf T})=w$, so that under \eqref{gauge-t}: ${\bf T}\rightarrow\Omega^w{\bf T}$. Then, the Weyl gauge differential of the tensor, its Weyl gauge derivative and Weyl gauge covariant derivative, respectively, are defined as it follows:

\bea &&d^*{\bf T}:=d{\bf T}+\frac{w}{2}Q^*_\lambda dx^\lambda{\bf T},\nonumber\\
&&\der^*_\alpha{\bf T}:=\der_\alpha{\bf T}+\frac{w}{2}Q^*_\alpha{\bf T},\nonumber\\
&&\nabla^*_\alpha:=\nabla_\alpha+\frac{w}{2}Q^*_\alpha,\label{gauge-der}\eea where

\bea d^*{\bf T}=dx^\mu\der^*_\mu{\bf T},\label{rel-d-dm}\eea and

\bea Q^*_\alpha\equiv\frac{a}{s}Q_\alpha+\frac{b}{4s}Q_{\alpha\;\;\mu}^{\;\;\mu},\label{qaster}\eea is a linear combination of contributions $Q_\alpha\equiv Q^\mu_{\;\;\mu\alpha}$ and $Q_{\alpha\;\;\mu}^{\;\;\mu}$, with arbitrary constants $a$, $b$ and $s=a+b$. Notice that in $\tilde W_4$, where $Q_{\alpha\mu\nu}=Q_\alpha g_{\mu\nu}$, we have that $Q^*_\alpha$ coincides with the Weyl gauge vector: $Q^*_\alpha=Q_\alpha$.

The above definitions warrant that the gauge differential, the gauge derivative and the gauge covariant derivative, transform like the geometrical object itself, i. e., under \eqref{gauge-t}: 

\bea &&d^*{\bf T}\rightarrow\Omega^wd^*{\bf T},\nonumber\\
&&\der^*_\alpha{\bf T}\rightarrow\Omega^w\der^*_\alpha{\bf T},\nonumber\\
&&\nabla^*_\alpha{\bf T}\rightarrow\Omega^w\nabla^*_\alpha{\bf T}.\nonumber\eea 

Since it will be useful in the subsequent analysis, as an illustration, let us write the gauge covariant derivative of the metric tensor (the conformal weight of the metric $w({\bf g})=2$):

\bea \nabla^*_\alpha g_{\mu\nu}=-Q_{\alpha\mu\nu}+Q^*_\alpha g_{\mu\nu}.\label{gcov-der-g}\eea Notice that in $\tilde W_4$, since $Q_{\alpha\mu\nu}=Q_\alpha g_{\mu\nu}$ and $Q^*_\alpha=Q_\alpha$, the gauge covariant derivative of the metric vanishes: $\nabla^*_\alpha g_{\mu\nu}=0$.


\subsection{Parallel transport in $\tilde W_4$ space}

In order to discuss on the notion of gauge invariant parallel transport in nonmetricity spaces, we start focusing the discussion in standard Weyl space $\tilde W_4$, where the analysis is simpler and then we shall aim at its generalization to $W_4$. 

Let ${\cal C}$ be a curve in $\tilde W_4$ that is parametrized by the affine parameter $\xi$: $x^\mu(\xi)$. We can define the gauge covariant derivative along the path $x^\mu(\xi)$ to be given by the following operator:

\bea \frac{D^*}{d\xi}:=\frac{dx^\mu}{d\xi}\nabla^*_\mu,\label{gcov-der-path}\eea where the gauge covariant derivative $\nabla^*_\mu$ is given by \eqref{gauge-der} with $Q^*_\alpha=Q_\alpha$ -- the Weyl gauge vector. Then, the parallel transport of given tensor ${\bf T}$ with coordinate components $T^{\alpha_1\alpha_2\cdots\alpha_i}_{\beta_1\beta_2\cdots\beta_j}$, along the path $x^\mu(\xi)$, is defined by the following requirement (this definition coincides with the one in \cite{dirac-1973, maeder-1978}):

\bea \frac{D^*{\bf T}}{d\xi}:=\frac{dx^\mu}{d\xi}\nabla^*_\mu{\bf T}=0\;\Leftrightarrow\;\frac{D^*}{d\xi}T^{\alpha_1\alpha_2\cdots\alpha_i}_{\beta_1\beta_2\cdots\beta_j}=0.\label{gcov-der-path-t}\eea 

Let us show that \eqref{gcov-der-path-t} is valid only in standard Weyl space $\tilde W_4$. For this purpose let us consider the parallel transport law \eqref{gcov-der-path-t} applied to a space-like unit vector ${\bf t}$ along the path ${\cal C}$ in generalized Weyl space $W_4$ (arbitrary nonmetricity), with coordinates $x^\mu(\xi)$. The unit vector has coordinate components $t^\mu$, such that:

\bea ({\bf t},{\bf t})=g_{\mu\nu}t^\mu t^\nu=1,\label{unit-t}\eea and its weight is $w({\bf t})=-1$. Applying \eqref{gcov-der-path} to both sides of \eqref{unit-t} and taking into account \eqref{gcov-der-g}, we get that:

\bea \frac{D^*t^\alpha}{d\xi}=J^\alpha_{\;\;\mu\nu}\frac{dx^\mu}{d\xi}\,t^\nu,\label{cov-der-t}\eea where for compactness of writing we have introduced the gauge invariant tensor:\footnote{Since under the gauge transformations \eqref{gauge-t} $Q^*_\alpha$ transforms like $Q_\alpha$, the tensor $J^\alpha_{\;\;\mu\nu}$ is gauge invariant.}

\bea &&J^\alpha_{\;\;\mu\nu}:=\frac{1}{2}\left(Q^{\;\;\alpha}_{\mu\;\;\nu}-Q^*_\mu\,\delta^\alpha_\nu\right),\nonumber\\
&&J_{\mu\alpha\nu}=\frac{1}{2}\left(Q_{\alpha\mu\nu}-Q^*_\alpha g_{\mu\nu}\right).\label{useful-t}\eea Notice that this tensor is symmetric in its first and third indices: $J_{\mu\alpha\nu}=J_{\nu\alpha\mu}$.

Hence, in generalized Weyl space $W_4$ with arbitrary nonmetricity $Q_{\alpha\mu\nu}\neq Q_\alpha g_{\mu\nu}$, the unit vector ${\bf t}$ is not parallel transported along ${\cal C}$ in the sense of \eqref{gcov-der-path-t}. Only in the particular subclass of standard Weyl space $\tilde W_4\subset W_4$, since $Q_{\alpha\mu\nu}=Q_\alpha g_{\mu\nu}$, the tensor \eqref{useful-t} vanishes and the unit vector is parallel transported according to \eqref{gcov-der-path-t}. From equations \eqref{gcov-der-g}, \eqref{unit-t} and \eqref{cov-der-t} it follows that:

\bea \frac{D^*t_\alpha}{d\xi}=\frac{dx^\mu}{d\xi}\nabla_\mu^*t_\alpha=-J^\nu_{\;\;\mu\alpha}\frac{dx^\mu}{d\xi}t_\nu.\label{cov-der-t'}\eea


\subsection{Dagger derivative}


It is convenient to the define the gauge covariant ``dagger'' derivative (or just dagger derivative for short) of the unit tangent vector with components $t^\alpha$:

\bea \nabla^\dag_\alpha t^\beta:=\nabla_\alpha^*t^\beta-J^\beta_{\;\;\alpha\mu}t^\mu,\label{dag-der-unit-vec}\eea and

\bea \nabla^\dag_\alpha t_\beta:=\nabla^*_\alpha t_\beta+J^\mu_{\;\;\alpha\beta}t_\mu.\label{dag-der-unit-vec'}\eea In terms of the dagger derivative equations \eqref{cov-der-t} and \eqref{cov-der-t'} can be written in the form of parallel transport equations along the worldline $x^\alpha(\xi)$:

\bea &&\frac{D^\dag t^\alpha}{d\xi}:=\frac{dx^\mu}{d\xi}\nabla^\dag_\mu t^\alpha=0,\nonumber\\
&&\frac{D^\dag t_\alpha}{d\xi}:=\frac{dx^\mu}{d\xi}\nabla^\dag_\mu t_\alpha=0,\label{pte-unit-vec}\eea respectively.

For an arbitrary vector ${\bf v}$ with coordinate components $v^\alpha$ and conformal weight $w({\bf v})=w$, its dagger derivative is defined as it follows:

\bea \nabla^\dag_\alpha v^\beta:=\nabla_\alpha^*v^\beta+wJ^\beta_{\;\;\alpha\mu}v^\mu,\label{gauge-der-v}\eea while

\bea \nabla^\dag_\alpha v_\beta:=\nabla^*_\alpha v_\beta+\bar wJ^\mu_{\;\;\alpha\beta}v_\mu,\label{gauge-der-v'}\eea where $w(v_\alpha)=\bar w=2+w$ is the conformal weight of the covariant components of the vector ${\bf v}$.

In general, for an arbitrary $(p,q)$-tensor ${\bf T}$, with coordinate components $T^{\alpha_1\alpha_2\cdots\alpha_p}_{\beta_1\beta_2\cdots\beta_q}$ and with conformal weight $w({\bf T})=w$, its dagger derivative is defined as:

\bea &&\nabla_\mu^\dag T^{\alpha_1\alpha_2\cdots\alpha_p}_{\beta_1\beta_2\cdots\beta_q}=\nabla_\mu^*T^{\alpha_1\alpha_2\cdots\alpha_p}_{\beta_1\beta_2\cdots\beta_q}+\frac{w}{p+q}\left(J^{\alpha_1}_{\mu\nu}T^{\nu\alpha_2\cdots\alpha_p}_{\beta_1\beta_2\cdots\beta_q}\right.\nonumber\\
&&\left.\;\;\;\;\;\;\;\;\;\;\;\;\;\;\;\;\;\;\;\;\;\;\;+\cdots+J^{\alpha_p}_{\mu\nu}T^{\alpha_1\alpha_2\cdots\nu}_{\beta_1\beta_2\cdots\beta_q}+J^\nu_{\mu\beta_1}T^{\alpha_1\alpha_2\cdots\alpha_p}_{\nu\beta_2\cdots\beta_q}\right.\nonumber\\
&&\left.\;\;\;\;\;\;\;\;\;\;\;\;\;\;\;\;\;\;\;\;\;\;\;+\cdots+J^\nu_{\mu\beta_q}T^{\alpha_1\alpha_2\cdots\alpha_p}_{\beta_1\beta_2\cdots\nu}\right).\label{dag-der-t}\eea


\subsection{Parallel transport law in $W_4$ space}

In order to consistently define parallel transport in $W_4$, in a such a way that any vector or tensor can be parallel transported in the usual way: constant coordinate components along parallel transport curve, we must introduce what we call as dagger derivative along the path of parallel transport $\cal C$ with coordinates $x^\alpha(\xi)$. 

Consider an arbitrary vector ${\bf v}$ with coordinate components $v^\alpha$ and conformal weight $w({\bf v})=w$. The dagger derivative of this vector along $\cal C$ is defined in the following way:

\bea \frac{D^\dag v^\alpha}{d\xi}:=\frac{dx^\mu}{d\xi}\nabla_\mu^\dag v^\alpha.\label{dag-der-v}\eea We require that under an infinitesimal parallel transport along the curve ${\cal C}$, the components of this vector do not change:

\bea &&\frac{D^\dag v^\alpha}{d\xi}=0.\nonumber\eea 

In general, for an arbitrary $(p,q)$-tensor ${\bf T}$, with coordinate components $T^{\alpha_1\alpha_2\cdots\alpha_p}_{\beta_1\beta_2\cdots\beta_q}$ and with conformal weight $w({\bf T})=w$, its dagger derivative is given by \eqref{dag-der-t}. Hence, the parallel transport law requires that the dagger derivative of the tensor ${\bf T}$ along the path of parallel transport $\cal C$, vanishes:

\bea \frac{D^\dag{\bf T}}{d\xi}:=\frac{dx^\mu}{d\xi}\nabla^\dag_\mu{\bf T}=0.\label{ptl}\eea This equation drives parallel transport of tensor ${\bf T}$ along $x^\alpha(\xi)$ in generalized Weyl space $W_4$. Worth noting that in standard Weyl space $\tilde W_4$, since $Q_{\alpha\mu\nu}=Q_\alpha g_{\mu\nu}$, then $J^\alpha_{\;\;\mu\nu}=0$, so that $\nabla^\dag_\alpha\rightarrow\nabla^*_\alpha$, $D^\dag/d\xi\rightarrow D^*/d\xi$, and \eqref{ptl} transforms into \eqref{gcov-der-path-t}.

Of course, different statements of the parallel transport law lead to different results. In \cite{my-arxiv-paper}, for instance, the parallel transport law \eqref{gcov-der-path-t} is applied to generalized Weyl space $W_4$, which led to the conclusion that unit tangent vectors with weight $w=-1$ do not obey the parallel transport equation \eqref{gcov-der-path-t} in $W_4$. What we have shown in this section is that it is possible to postulate the parallel transport law \eqref{ptl}, which is satisfied by any vectors and tensors in $W_4$ space.\footnote{A similar situation to the one described above occurs in \cite{2clock-tomi} during the demonstration of lemma 2, where the assumed statement of parallel transport law is not satisfied by unit tangent vectors of weight $w=-1$ (see section \ref{sect-challenge}).}


\subsection{Variation of length during parallel transport}

It has been well established in the bibliography that in Weyl geometry spaces, during parallel transport, the length of vectors (and tensors) varies and depends on followed path. Given that there has been renaissance of old discussions about this issue \cite{2clock-tomi, 2c-delhom, hobson, my-arxiv-paper, hobson-replay}, in this section, for sake of generality, we aim at explaining this effect in spaces with arbitrary nonmetricity $Q_{\alpha\mu\nu}$.

Let us assume that the vector ${\bf v}$ with components $v^\alpha$ and conformal weight $w({\bf v})=w$, is submitted to parallel transport along the path ${\cal C}$, which, as before, is parametrized by the affine parameter $\xi$: $x^\mu=x^\mu(\xi)$. The length $v\equiv||{\bf v}||$ of the vector is defined as,

\bea v^2=g_{\mu\nu}v^\mu v^\nu,\label{length}\eea where for space-like vector $v^2>0$, while for timelike vector $v^2<0$ ($v$ is an imaginary quantity). Taking the gauge covariant derivative along the path ${\cal C}$ in both sides of \eqref{length}, we obtain that,

\bea \frac{D^*v^2}{d\xi}=\frac{D^*g_{\mu\nu}}{d\xi}v^\mu v^\nu+2g_{\mu\nu}v^\nu\frac{D^*v^\mu}{d\xi}.\label{eq1}\eea If realize that along the path of parallel transport:

\bea &&\frac{D^*g_{\mu\nu}}{d\xi}=-2J_{\mu\lambda\nu}\frac{dx^\lambda}{d\xi},\label{eq2}\eea and if further take into account \eqref{gauge-der-v}, then equation \eqref{eq1} can be written in the form of the parallel transport law \eqref{ptl}:

\bea \frac{D^\dag v^2}{d\xi}=\frac{dx^\lambda}{d\xi}\nabla^\dag_\lambda v^2=0,\label{ptl-scalar}\eea where we have defined the dagger covariant derivative of the length squared

\bea \nabla^\dag_\alpha v^2:=\nabla^*_\alpha v^2+2(1+w)J_{\mu\lambda\nu}v^\mu v^\nu.\label{dag-der-scalar}\eea Besides, if define the spacelike unit vector ${\bf t}$:

\bea {\bf t}:=\frac{\bf v}{v}\;\Leftrightarrow\;t^\alpha=\frac{v^\alpha}{v},\;({\bf t},{\bf t})=g_{\mu\nu}t^\mu t^\nu=1,\label{unit-v}\eea where the conformal weight $w({\bf t})=-1$, and recall the definition of gauge differential in \eqref{gauge-der}, then the parallel transport law \eqref{ptl-scalar} can be finally written in the form:

\bea \frac{D^\dag v^2}{d\xi}=0\;\Rightarrow\;\frac{d\ln v}{d\xi}=-\frac{w+1}{2}Q_{\lambda\mu\nu}t^\mu t^\nu\frac{dx^\lambda}{d\xi}.\label{eq3}\eea Formal integration of \eqref{eq3},

\bea \Delta\ln v=-\frac{w+1}{2}\int_{\cal C}Q_{\lambda\mu\nu}t^\mu t^\nu dx^\lambda,\label{var-ln}\eea leads to the equation of variation of the length of vector ${\bf v}$ during parallel transport along path ${\cal C}$, from the origin $x=0$ to the spacetime point $x$:

\bea v(x)=v(0)\exp{\left[-\frac{w+1}{2}\int_{\cal C}Q_{\lambda\mu\nu}t^\mu t^\nu dx^\mu\right]},\label{master-eq}\eea where $v(0)=C$ is an integration constant which we identify with the magnitude of the length of vector ${\bf v}$, evaluated at the starting point of the worldline $\cal C$. 

Let us apply equation \eqref{master-eq} to the four-momentum vector ${\bf p}$, which is at the core of the SCE. Let the path of the particle be parameterized by the proper time $\tau$. The coordinate components of the four-velocity vector ${\bf u}$ are $u^\mu:=dx^\mu/d\tau$, where $d\tau=ids$ ($s$ is the arc-length and the imaginary unit $i$ arises due to our metric signature choice). Consequently, the length of the four-velocity vector $u\equiv||{\bf u}||=\pm i$, while its conformal weight $w({\bf u})=-1$. After the above specifications we have that,

\bea {\bf p}:=m{\bf u}\;\Rightarrow\;p\equiv||{\bf p}||=\pm im,\label{4-p}\eea where $m$ is the mass of a point particle and the conformal weight of the four-momentum $w({\bf p})=-2$ (the conformal weight of the mass parameter $w(m)=-1$). If in \eqref{master-eq} we make the replacement ${\bf p}\rightarrow{\bf v}$, we obtain the following equation:

\bea m(x)=m(0)\exp{\left[-\frac{1}{2}\int_{\cal C}Q_{\lambda\mu\nu}u^\mu u^\nu dx^\lambda\right]},\label{sce-eq}\eea where we have taken into consideration that the components of the spacelike unit vector ${\bf t}$, that is tangent to ${\cal C}$, are given by:

\bea t^\mu=\frac{p^\mu}{p}=\frac{mdx^\mu/d\tau}{\pm im}=\pm i\frac{dx^\mu}{d\tau}=\pm iu^\mu.\label{t-p}\eea 

Equation \eqref{sce-eq} quantifies the variation of the mass of a particle which is moving with speed $u^\mu=dx^\mu/d\tau$ along the path ${\cal C}$. The dependence of the mass parameter on the speed pattern of the particle during its motion, which is apparent in \eqref{sce-eq}, is a new effect that does not arise in $\tilde W_4$ space.


\subsubsection{Standard Weyl space $\tilde W_4$}

In $\tilde W_4$, since $Q_{\alpha\mu\nu}=Q_\alpha g_{\mu\nu}$ and, consequently the vector $Q^*_\mu=Q_\mu$, coincides with the Weyl gauge vector, equation \eqref{master-eq} simplifies to:

\bea v(x)=v(0)\exp{\left(-\frac{w+1}{2}\int_{\cal C}Q_\mu dx^\mu\right)},\label{master-eq'}\eea while the equation \eqref{sce-eq} transforms into the following equation:

\bea m(x)=m(0)\exp{\left(\frac{1}{2}\int_{\cal C}Q_\mu dx^\mu\right)}.\label{sce-eq'}\eea 

Equation \eqref{sce-eq} is the basis for the explanation of the SCE in $W_4$, while \eqref{sce-eq'} drives the SCE in $\tilde W_4$.


\subsection{Variation of the inner product of vectors during parallel transport}\label{subsect-inner-p}

For generality, let us consider also the parallel transport of the inner product of two vectors. Take two vector fields ${\bf v}$ and $\bf w$ with coordinate components $v^\mu$, $w^\mu$ and conformal weights $w({\bf v})=w_{\bf v}$ and $w({\bf w})=w_{\bf w}$, respectively. Let these vector fields be parallel transported along the curve ${\cal C}$, that is parametrized by $\xi$: $D^\dag v^\mu/d\xi=0$ and $D^\dag w^\mu/d\xi=0$. Using the same procedure that we followed above, we get that:

\bea &&\frac{d({\bf v},{\bf w})}{d\xi}=-(2+w_{\bf v}+w_{\bf w})Q_{\lambda\mu\nu}\frac{dx^\lambda}{d\xi}v^\mu w^\nu.\label{app-1}\eea Next, take into account the following chain of equalities:\footnote{Recall that $v:=||{\bf v}||$ and $w:=||{\bf w}||$ are the lengths of vectors ${\bf v}$ and ${\bf w}$, respectively.}

\bea &&({\bf v},{\bf w})=vw\cos\theta,\;t^\mu_{\bf v}=\frac{v^\mu}{v},\;t^\mu_{\bf w}=\frac{w^\mu}{w}\nonumber\\
&&\Rightarrow\;v^\mu w^\mu=t^\mu_{\bf v}t^\nu_{\bf w}vw=\frac{1}{\kappa}t^\mu_{\bf v}t^\nu_{\bf w}({\bf v},{\bf w}),\label{app-2}\eea where $t^\mu_{\bf v}$and $t^\mu_{\bf w}$ are spacelike unit vectors, and $\kappa\equiv\cos\theta$ ($\theta$ is the angle between vectors $\bf v$ and $\bf w$) is a real constant taking values in the interval $[-1,1]$. If substitute \eqref{app-2} into \eqref{app-1}, we get that,

\bea \frac{d\ln({\bf v},{\bf w})}{d\xi}=-\frac{2+w_{\bf v}+w_{\bf w}}{\kappa}Q_{\lambda\mu\nu}t^\mu_{\bf v}t^\nu_{\bf w}\frac{dx^\lambda}{d\xi},\label{app-5}\eea whose integration along the closed path ${\cal C}$ yields to:

\bea \Delta\ln({\bf v},{\bf w})=-\frac{2+w_{\bf v}+w_{\bf w}}{\kappa}\int_{\cal C}Q_{\lambda\mu\nu}t^\mu_{\bf v}t^\nu_{\bf w}dx^\lambda.\label{app-6}\eea In the latter equation the quantity $\kappa$ has been taken out of the integral since, as it can be straightforwardly demonstrated, it does not depend on path:\footnote{We have that $\kappa=({\bf t}_{\bf v},{\bf t}_{\bf w})=g_{\mu\nu} t^\mu_{\bf v}t^\nu_{\bf w}$. Take the gauge covariant derivative along the path $x^\alpha(\xi)$, in both sides of this equation: $$\frac{d\kappa}{d\xi}=\frac{D^*g_{\mu\nu}}{d\xi}t^\mu_{\bf v}t^\nu_{\bf w}+g_{\mu\nu}\frac{D^*t^\mu_{\bf v}}{d\xi}t^\nu_{\bf w}+g_{\mu\nu}t^\mu_{\bf v}\frac{D^*t^\nu_{\bf w}}{d\xi}.$$ Hence, according to \eqref{cov-der-t}, we get that, $$\frac{d\kappa}{d\xi}=\frac{D^*g_{\mu\nu}}{d\xi}t^\mu_{\bf v}t^\nu_{\bf w}+\left(J_{\mu\alpha\nu}t^\mu_{\bf v}t^\nu_{\bf w}+J_{\nu\alpha\mu}t^\mu_{\bf v}t^\nu_{\bf w}\right)\frac{dx^\alpha}{d\xi},$$ so that, if take into account \eqref{eq2} and the symmetry of tensor $J_{\mu\alpha\nu}$ \eqref{useful-t} in its first and third indices, we finally obtain that $d\kappa/d\xi=0$.} $d\kappa/d\xi=0$. Equation \eqref{app-6} can be written in the following way:

\bea ({\bf v},{\bf w})=({\bf v},{\bf w})_0\;e^{-\frac{2+w_{\bf v}+w_{\bf w}}{\kappa}\int_{\cal C}Q_{\lambda\mu\nu}t^\mu_{\bf v}t^\nu_{\bf w}dx^\lambda}.\label{app-6'}\eea 

It should be emphasized that equations \eqref{gcov-der-g}, \eqref{gcov-der-path}, \eqref{gcov-der-path-t}, \eqref{cov-der-t}, \eqref{gauge-der-v}, \eqref{gauge-der-v'}, \eqref{dag-der-v} are a direct consequence of the nonmetricity law \eqref{gen-nm} of generalized Weyl space $W_4$, while \eqref{ptl} is our assumed parallel transport law. Equations \eqref{var-ln}, \eqref{master-eq}, \eqref{sce-eq} and related equations \eqref{app-6} and \eqref{app-6'} above, are also consequence of the nonmetricity law \eqref{gen-nm} and of \eqref{ptl}. This means that if \eqref{gen-nm} is valid in $W_4$ for arbitrary nonmetricity $Q_{\alpha\mu\nu}$ and the parallel transport law \eqref{ptl} is assumed, the mentioned and related equations are valid in $W_4$ as well. This conclusion is important because the adoption of $W_4$ as the underlying geometric background space, entails that the SCE is inevitable. Otherwise, if one denies the physical occurrence of the SCE, one renounces to equations like \eqref{sce-eq} and to the nonmetricity law \eqref{gen-nm} and also to the parallel transport law \eqref{ptl}, i. e., one renounces to identify physical vectors and tensors with corresponding hypothetical vectors and tensors in $W_4$. Consequently, one gives up the description of physical phenomena in $W_4$ space (see the related discussion in \cite{my-arxiv-paper}).



\section{Gauge symmetry, autoparallels and geodesics}\label{sect-geod}


In general autoparallels -- ``straightest curves'' of the geometry -- do not coincide with the geodesics, which are the ``shortest curves'' \cite{poplawski-arxiv, adak-arxiv, obukhov}. There goes a discussion on whether autoparallels or geodesics describe the motion of test particles \cite{adak-arxiv, obukhov}. However, there are particular cases when autoparallels and geodesics coincide as, for instance, in GR. As we shall see, these coincide as well in $\tilde W_4$. Anyway, geodesics and autoparalles can be associated exclusively with the motion of spinless test (point) particles. Spinor fields like the fermions obey the Dirac equation in curved background, while extended spinning test bodies obey the Mathisson-Papapetrou-Dixon equations \cite{mathisson, papapetrou, dixon, wald} (see section \ref{sect-fermions}).

Here we obtain a bit different results compared with those that were obtained in \cite{my-arxiv-paper}, on the basis of a different transport law \eqref{gcov-der-path-t} instead of \eqref{ptl}.


\subsection{Auto-parallels}

In generalized Weyl space $W_4$ the ``timelike'' autoparallels are those curves along which the gauge covariant derivative of the tangent four-velocity vector ${\bf u}$, vanishes. Here $u^\mu=dx^\mu/d\tau$ are the coordinate components of ${\bf u}$ and, as long as this does not cause loss of generality, we chose the proper time $\tau$ to be the affine parameter along the autoparallel curve. The conformal weight of the four-velocity vector $w({\bf u})=-1$. In other words, the autoparallel curves satisfy:

\bea &&\frac{D^\dag u^\alpha}{d\tau}=u^\mu\nabla^\dag_\mu u^\alpha=0\;\Rightarrow\nonumber\\
&&\frac{du^\alpha}{d\tau}+\Gamma^\alpha_{\;\;\mu\nu}u^\mu u^\nu-\frac{1}{2}Q^*_\mu u^\mu u^\alpha-J^\alpha_{\;\;\mu\nu}u^\mu u^\nu=0,\nonumber\eea or, in explicit form, in terms of the arc-length $d\tau\rightarrow ids$:

\bea &&\frac{d^2x^\alpha}{ds^2}+\Gamma^\alpha_{\;\;\mu\nu}\frac{dx^\mu}{ds}\frac{dx^\nu}{ds}-\frac{1}{2}Q^{\;\;\alpha}_{\mu\;\;\nu}\frac{dx^\mu}{ds}\frac{dx^\nu}{ds}=0\;\Leftrightarrow\nonumber\\
&&\frac{d^2x^\alpha}{ds^2}+\{^\alpha_{\mu\nu}\}\frac{dx^\mu}{ds}\frac{dx^\nu}{ds}\nonumber\\
&&\;\;\;\;\;\;\;\;\;\;+\frac{1}{2}\left(Q^{\;\;\alpha}_{\mu\;\;\nu}-Q^\alpha_{\;\;\mu\nu}\right)\frac{dx^\mu}{ds}\frac{dx^\nu}{ds}=0.\label{time-auto-p}\eea 

In the same fashion, in $W_4$ the ``null'' autoparallels are those curves along which the gauge covariant derivative of the wave vector ${\bf k}$ with components $k^\mu:=dx^\mu/d\lambda$ ($\lambda$ is a parameter along the null autoparallel), vanishes:

\bea &&\frac{D^\dag k^\alpha}{d\lambda}=k^\mu\nabla^\dag_\mu k^\alpha=0\;\Rightarrow\nonumber\\
&&\frac{dk^\alpha}{d\lambda}+\Gamma^\alpha_{\;\;\mu\nu}k^\mu k^\nu-Q^*_\mu k^\mu k^\alpha-2J^\alpha_{\;\;\mu\nu}k^\mu k^\nu=0,\nonumber\eea where we have taken into account that the conformal weight of the wave vector $w({\bf k})=-2$, i. e., it coincides with the weight of the four-momentum since, in the quantum limit both should be related by ${\bf p}=\hbar{\bf k}$.\footnote{Here it is implicitly assumed that the universal constant $\hbar$ is not transformed by the Weyl gauge transformations.} Then, the autoparallel null curves satisfy the following equations:

\bea &&\frac{dk^\alpha}{d\lambda}+\Gamma^\alpha_{\;\;\mu\nu}k^\mu k^\nu-Q^{\;\;\alpha}_{\mu\;\;\nu} k^\mu k^\nu=0\;\Leftrightarrow\nonumber\\
&&\frac{dk^\alpha}{d\lambda}+\{^\alpha_{\mu\nu}\}k^\mu k^\nu-\frac{1}{2}Q^\alpha_{\;\;\mu\nu}k^\mu k^\nu=0.\label{null-auto-p}\eea

In standard Weyl space $\tilde W_4$, since $Q_{\alpha\mu\nu}=Q_\alpha g_{\mu\nu}$, then, according to \eqref{qaster},

\bea Q^*_\alpha=\frac{a}{s}Q_\alpha+\frac{b}{4}Q_\alpha\delta^\mu_\mu=\frac{a+b}{s}Q_\alpha=Q_\alpha.\nonumber\eea In this case for the timelike autoparallels we have that:

\bea \frac{d^2x^\alpha}{ds^2}+\{^\alpha_{\mu\nu}\}\frac{dx^\mu}{ds}\frac{dx^\nu}{ds}-\frac{1}{2}Q_\mu h^{\mu\alpha}=0,\label{time-like-autop-sws}\eea where

\bea h^{\mu\alpha}:=g^{\mu\alpha}+u^\mu u^\alpha=g^{\mu\alpha}-\frac{dx^\mu}{ds}\frac{dx^\alpha}{ds},\label{orto-proj}\eea is the orthogonal projection tensor, which projects any vector or tensor onto the hypersurface orthogonal to the four-velocity vector $u^\mu=dx^\mu/d\tau$. Meanwhile, for null autoparallels, since $$Q^\alpha_{\;\;\mu\nu}k^\mu k^\nu=Q^\alpha g_{\mu\nu}k^\mu k^\nu=0,$$ one obtains the standard GR result:

\bea \frac{dk^\alpha}{d\lambda}+\{^\alpha_{\mu\nu}\}k^\mu k^\nu=0.\label{null-geod}\eea


\subsection{Geodesic equations}

The geodesic equations are equations of motion in the sense that these are the result of applying the variational principle of least action. Time-like and null particles follow geodesics. When these are compared with the corresponding auto-parallels one can measure how much the motion paths depart from the straightest curves of the geometry.

In the GR context the action of timelike particles reads $S=m\int ds$, where $m$ is the constant mass parameter. In generalized Weyl space $W_4$, since the mass, being the squared length of the four-momentum of the particle, varies in spacetime, then $m$ can not be taken out of the action integral. The action integral in $W_4$ reads:

\bea S=\int mds.\nonumber\eea From this action the following equations of motion -- geodesic equations -- can be derived (see \cite{my-arxiv-paper} for details of the derivation):

\bea \frac{d^2x^\alpha}{ds^2}+\{^\alpha_{\mu\nu}\}\frac{dx^\mu}{ds}\frac{dx^\nu}{ds}-\frac{1}{m}\frac{\delta m}{\delta x^\mu}h^{\mu\alpha}=0,\label{time-geod'}\eea where the non-Riemannian term $\propto\delta m/m\delta x^\mu$ accounts for the variation of mass during parallel transport. If one writes equation \eqref{sce-eq} in variational form, one obtains:\footnote{Notice that we keep using variation instead of differentiation to underline that, in general, $\delta m$ is not a perfect differential.} 

\bea \frac{1}{m}\frac{\delta m}{\delta x^\alpha}=-\frac{1}{2}Q_{\alpha\mu\nu} u^\mu u^\nu=\frac{1}{2}Q_{\alpha\mu\nu}\frac{dx^\mu}{ds}\frac{dx^\nu}{ds}.\label{dm-eq}\eea 

In order to determine the term $\propto\delta m/m\delta x^\alpha$ in \eqref{time-geod'} we assume a ``consistency'' hypothesis \cite{my-arxiv-paper}. This hypothesis (or postulate if you want) allows to identify hypothetical vectors and tensors living in $W_4$ (for instance the four-momentum \eqref{4-p}) with the corresponding physical vectors and tensors. Otherwise one can not describe given physical phenomenon in $W_4$ space. In the present case the most reasonable consistency hypothesis is to identify the physical mass parameter in \eqref{time-geod'} with the hypothetical mass parameter $m=ip$ which obeys \eqref{sce-eq}. In other words, we substitute equation \eqref{dm-eq} back into \eqref{time-geod'}. We get that:

\bea \frac{d^2x^\alpha}{ds^2}+\{^\alpha_{\mu\nu}\}\frac{dx^\mu}{ds}\frac{dx^\nu}{ds}-\frac{h^{\alpha\mu}}{2}Q_{\mu\lambda\nu}\frac{dx^\lambda}{ds}\frac{dx^\nu}{ds}=0.\label{time-like-geod}\eea Worth noting that, in standard Weyl space $\tilde W_4$, since $Q_{\alpha\mu\nu}=Q_\alpha g_{\mu\nu}$, equation \eqref{time-like-geod} transforms into the timelike autoparallel \eqref{time-like-autop-sws} of $\tilde W_4$. Hence, the autoparallels of $\tilde W_4$ and the corresponding geodesics, coincide. 

This is true as well of the null autoparallels and geodesics. Actually, the null geodesic equations can be derived from the following action:

\bea S_\text{null}=\frac{1}{2}\int g_{\mu\nu}\dot x^\mu\dot x^\nu d\xi,\label{action-null-geod}\eea where the dot accounts for derivative with respect to the parameter $\lambda$ of the path $x^\mu(\lambda)$ followed by photons (by radiation in general). From \eqref{action-null-geod} the GR null geodesic equations \eqref{null-geod} are obtained. These coincide with the null autoparallels. 

The null geodesic equations do not depend on $Q_{\alpha\mu\nu}$. This means that photons and radiation probe the Riemann affine structure of spacetime. In other words, photons and radiation interact only with the metric field, i. e., with the LC curvature of spacetime. These do not interact with nonmetricity.



\section{Gauge symmetry and the second clock effect}\label{sect-sce} 


Let us base the physical analysis of the SCE on the functioning of an atomic clock which measures the International Atomic Time. The principle of operation of an atomic clock is based on atomic physics: it measures the electromagnetic signal that electrons in atoms emit when they change energy levels. For instance, the energy of each energy level in the hydrogen atom, labeled by $n$, is given by: $E_n\approx-m\alpha^2/2n^2$, where $m$ is the mass of the electron and $\alpha\approx 1/137$ is the fine-structure constant. Any changes in the mass $m$ over spacetime will cause changes in the energy levels and, consequently, in the energy of the atomic transitions $$\omega_{if}=|E_{n_f}-E_{n_i}|=\frac{m\alpha^2}{2}\left(\frac{1}{n^2_f}-\frac{1}{n^2_i}\right).$$ Hence, the functioning of atomic clocks will be affected by the variation of masses over spacetime, according to equation \eqref{sce-eq}. 

Let us consider a collection of identical atoms that are parallel transported along neighboring paths from the origin $x=0$ to a given point $x$. Let us take the larger difference arising between the masses of any two atoms in the collection at $x$: $\Delta m=m(x)-\bar m(x)$. Then, according to \eqref{sce-eq} one gets the following gauge invariant ratio: 

\bea \frac{\Delta\omega_{if}}{\omega_{if}}=\frac{\Delta m}{m}=1-\exp{\left[Q^{{\bf u}'}_{{\cal C}'}(x)-Q^{\bf u}_{\cal C}(x)\right]},\label{g-ratio}\eea where $\Delta\omega_{if}$ quantifies the broadening of the given spectral line and we have adopted the following notation (recall that $u^\mu=dx^\mu/d\tau$ is the four-velocity):

\bea Q^{\bf u}_{\cal C}(x)\equiv-\frac{1}{2}\int_{\cal C}Q_{\mu\sigma\nu} u^\sigma u^\nu dx^\mu.\label{quc}\eea 

The novel feature in $W_4$ is that the broadening of the spectral lines depends not only on the followed path, but also on the speed pattern. Hence, even if two identical atoms at the origin, are parallel transported along a same path to the distant point $x$, but following different speed patterns, at $x$ the emission/absorption lines of one of the atoms will be shifted with respect to the same spectral lines of the other atom. In order to illustrate this novel feature, let us for brevity make the following identification: 

\bea &&\Delta Q^{{\bf u}'{\bf u}}_{\cal C}\equiv Q^{{\bf u}'}_{\cal C}(x)-Q^{\bf u}_{\cal C}(x)\nonumber\\
&&\;\;\;\;\;\;\;\;\;\;\;=\frac{1}{2}\int_{\cal C}Q_{\lambda\mu\nu}\left(u^\mu u^\nu-u'^\mu u'^\nu\right)dx^\lambda,\label{dquc}\eea where $Q^{{\bf u}'}_{\cal C}$ and $Q^{\bf u}_{\cal C}$ are given by \eqref{quc}. In the above equation ${\bf u}'$ stands for the four-velocity of one of the atoms, while ${\bf u}$ represents the four-velocity vector of the second atom. Then, as long as ${\bf u}'\neq{\bf u}$, there is a non-vanishing relative shift of given spectral lines:

\bea \frac{\Delta\omega_{if}}{\omega_{if}}=1-\exp{\Delta Q^{{\bf u}'{\bf u}}_{\cal C}}.\label{u-broad}\eea 

When instead of just two atoms, a sample of atoms of given substance is considered, the above discussed shift leads to an effective broadening of spectral lines that may be quantified by the largest possible shift $\Delta\omega_{if}$. This shift is not to be confused with the GR shift of frequencies which is due to the propagation of photons in a curved background space, and is the same for any frequency. In Weyl space $W_4$ the first clock effect arises as well due to the Riemann-Christoffel (Levi-Civita) curvature of space, leading to the same shift of frequencies for any spectral line.\footnote{As seen in section \ref{sect-geod} photons and radiation interact only with the LC curvature of spacetime.} The SCE, on the contrary, is due to the above described shift of frequencies which is different for different frequencies: $\Delta\omega_{if}\propto\omega_{if}$, as it can be seen from equations \eqref{g-ratio} and \eqref{u-broad}.

In $\tilde W_4$ for the gauge-invariant ratio \eqref{g-ratio}, according to \eqref{sce-eq'}, one gets: 

\bea \frac{\Delta\omega_{if}}{\omega_{if}}=1-\exp{\frac{1}{2}\left(\int_{{\cal C}'}Q_\mu dx^\mu-\int_{\cal C}Q_\mu dx^\mu\right)}.\label{g-ratio'}\eea Notice that in this case the given spectral line is sharp: $\Delta\omega_{if}=0$, only if either ${\cal C}'={\cal C}$, or if $Q_\mu=\der_\mu\phi$, where $\phi$ is the Weyl gauge scalar. In this last case $\int_{\cal C}\der_\mu\phi dx^\mu=\phi(x)-\phi(0)$, independent of the path joining the starting and final points. WIG is the resulting geometric structure.



\section{Challenging the SCE}\label{sect-challenge}


There are in the bibliography points of view that are contrary to the occurrence of the SCE \cite{2clock-tomi, hobson, hobson-replay, adak-arxiv}. Let us first comment on \cite{2clock-tomi} and on \cite{adak-arxiv} which deal with the SCE in generalized Weyl space $W_4$ (specifically in its subsect $Z_4$), and then we shall discuss on the point of view developed in \cite{hobson, hobson-replay} where the SCE in $\tilde W_4$ is challenged. 

In \cite{2clock-tomi} it is demonstrated that the SCE does not arise in symmetric teleparallel theories.\footnote{Although in \cite{2clock-tomi} the torsion contribution is considered, here we omit it for simplicity and because it is behind the scope of the present paper.} This is done through lemma 2 of the mentioned reference. Their result is obtained by ignoring the gauge symmetry which is a manifest symmetry of generalized Weyl space $W_4$. Besides, among the most important assumptions in the demonstration of lemma 2, is that the covariant derivative of vectors fields -- take, for instance, the four-velocity vector ${\bf u}$ -- vanishes during parallel transport along a given curve $x^\mu(\xi)$: $\nabla_\mu u^\alpha=0$. Hence, even if renounce to gauge symmetry, the demonstration is not valid for tangent vectors of weight $w=-1$. Actually, consider a tangent vector field ${\bf u}$ to the curve ${\cal C}$ that is parametrized by $\xi$. Its conformal weight is $w({\bf u})=-1$, while its length $g_{\mu\nu}u^\mu u^\nu=-1$. By taking the covariant derivative of this last equation one gets: $$(\nabla_\alpha g_{\mu\nu})u^\mu u^\nu+2g_{\mu\nu}u^\nu\nabla_\alpha u^\mu=0,$$ from where it follows that: 

\bea \nabla_\alpha u^\mu=\frac{1}{2}Q^{\;\;\mu}_{\alpha\;\;\nu}u^\nu.\label{correct-der}\eea This result is valid for any tangent vector whose weight $w=-1$, so that its length has vanishing weight. This means that the result holds true even if consider gauge symmetry. 

The demonstration of lemma 2 in \cite{2clock-tomi} starts with the following equation for the variation of the inner product of two vectors ${\bf u}$ and ${\bf v}$ parallel transported along a closed path $\bar{\cal C}$ (equation (12) of the mentioned reference):

\bea \Delta\left({\bf u},{\bf v}\right)=-\oint_{\bar{\cal C}}Q_{\mu\alpha\beta}u^\alpha v^\beta dx^\mu.\label{12}\eea After applying Stoke's theorem we have that

\bea &&\oint_{\bar{\cal C}}Q_{\mu\alpha\beta}u^\alpha v^\beta dx^\mu=\nonumber\\
&&\iint_S\left[\nabla_{[\mu}Q_{\nu]\alpha\beta}u^\alpha v^\beta+\nabla_{[\mu}\left(u^\alpha v^\beta\right)Q_{\nu]\alpha\beta}\right]dx^\mu\wedge dx^\nu\nonumber\\
&&=\iint_S\left[R_{(\alpha\beta)\mu\nu}u^\alpha v^\beta+\nabla_{[\mu}\left(u^\alpha v^\beta\right)Q_{\nu]\alpha\beta}\right]dx^\mu\wedge dx^\nu,\nonumber\eea where, in the last line we have taken into account the third Bianchi identity \eqref{3-bianchi-id}. Then in \cite{2clock-tomi} it is assumed that, since vectors $\bf u$ and $\bf v$ are both parallel transported (their inner product to be specific), then both $\nabla_\mu u^\alpha=0$ and $\nabla_\mu v^\beta=0$, and the second term within the surface integral vanishes. As we shall see, this assumption is correct only if, besides ignoring gauge symmetry, none of the vectors ${\bf u}$ and ${\bf v}$ had weight $w=-1$. On the contrary, if one of the vectors in the inner product, say vector $\bf u$, has weight $w({\bf u})=-1$, then equation \eqref{correct-der} is to be satisfied. In this case

\bea &&\oint_{\bar{\cal C}}Q_{\mu\alpha\beta}u^\alpha v^\beta dx^\mu=\nonumber\\
&&\iint_S\left[R_{(\alpha\beta)\mu\nu}+\frac{1}{2}Q^{\;\;\lambda}_{\mu\;\;(\alpha}Q_{\nu\lambda\beta)}\right]u^\alpha v^\beta dx^\mu\wedge dx^\nu.\nonumber\eea This means that the demonstration of lemma 2 in \cite{2clock-tomi} is valid only if neither the weight of vector ${\bf u}$ nor that of vector ${\bf v}$ equal $-1$. If one of the vectors has weight $w=-1$, then instead of equation (15) of that reference, one obtains (see appendix \ref{app-a} for a gauge invariant version of the demonstration):

\bea \Delta\left({\bf u},{\bf v}\right)=-\frac{1}{2}\iint_S Q^{\;\;\lambda}_{\mu\;\;(\alpha}Q_{\nu\lambda\beta)}\;u^\alpha v^\beta dx^\mu\wedge dx^\nu.\label{15-correct}\eea where we have set $R_{(\alpha\beta)\mu\nu}=0$ in order to satisfy the teleparallel condition. Hence, even under the teleparallel condition there is a net (nonvanishing) variation of the inner product of vectors during parallel transport in a closed path, contrary to the result of \cite{2clock-tomi}. Yet, our main argument against demonstrations of the kind found in \cite{2clock-tomi} -- and in many other bibliographic references, including textbooks like \cite{adler-book} -- is in the closed path $\bar{\cal C}$ required \eqref{12}. In section \ref{sect-ctc} we shall comment on this.

Let us underline that the demonstration of lemma 2 of \cite{2clock-tomi}, is made on the basis of equation \eqref{12} for parallel transport of a scalar product of vectors -- equation (12) of the mentioned reference -- which is not a gauge covariant equation (compare with the related gauge covariant expression \eqref{app-6}). In appendix \ref{app-a} we show how gauge symmetry modifies the above result.

In reference \cite{adak-arxiv} the authors proposed a novel prescription for the parallel transport of tangent vectors that allows keeping its length the same during parallel transport along a given curve. This may be an interesting possibility that does not contradict our result that the SCE is inevitable in spaces with arbitrary nonmetricity. As we have shown the length of tangent vectors with conformal weight $w=-1$ does not change during parallel transport. However, the length of vectors with weight $w\neq -1$, inevitably changes along the path of parallel transport. This is particularly true for the four-momentum vector ${\bf p}=m{\bf u}$, which is central to the explanation of the SCE. Additionally, the authors of \cite{adak-arxiv} do not pay attention to the gauge symmetry, which is our guiding principle.


In references \cite{hobson, hobson-replay} it is claimed that the non-occurrence of the SCE is generic in $\tilde W_4$. In order to show that it is so, an elaborate explanation is given, which is based on the assumption that there are vectors such as the four-velocity and four-momentum of an atom, that are not parallel transported along atom's world line even if it is in free fall \cite{hobson-replay}. The above scenario is physically implemented by assuming that certain scalar (compensator) field $\vphi$ of non-geometric origin, gives masses to point-particles $\propto\vphi(x)$. Statements like ``there are physical vectors that are not parallel transported along given world line'', represent a very strong hypothesis, affecting the geometrical description of involved phenomenon. This statement rules out, for instance, generalized Weyl geometry space $W_4$, as a potential arena for the geometrical description of given gravitational phenomena.

As it is discussed at the end of section \ref{sect-p-transp}, the adoption of $W_4$ (and its subclass $\tilde W_4$) as the underlying geometric background space entails that the SCE is inevitable. Otherwise, renouncing to identify hypothetical vectors in $W_4$, such as the four-momentum \eqref{4-p}, with the corresponding physical vectors, such as the four-momentum of an atom, means that one is renouncing to the description of physical phenomena in $W_4$ space (see the discussion on this issue in \cite{my-arxiv-paper}).



\section{Consistency hypothesis and matter coupling to nonmetricity}\label{sect-fermions}


There is yet another counterargument that, according to \cite{cheng-1988, cheng-arxiv}, allows avoiding the issue associated with the SCE in standard Weyl space $\tilde W_4$ (vectorial nonmetricity: $Q_{\alpha\mu\nu}=Q_\alpha g_{\mu\nu}$). Although in the mentioned bibliographic references the argument was demonstrated in $\tilde W_4$ space exclusively, it may be extended to generalized Weyl space $W_4$ as well \cite{delhom-coupling, tomi-replay}. 

The argument exposed in \cite{cheng-1988, cheng-arxiv} goes like this. The Lagrangian density of the fermion coupled with the gravitational field reads (below, for simplicity, we omit $SU(2)\otimes U(1)$ gauge terms):

\bea {\cal L}_\text{fermion}=i\bar{\psi}\cancel{\hat D}\psi,\label{psi-lag}\eea where $\psi$ is the Dirac spinor ($\bar{\psi}$ is its adjoint) and the slash gauge derivative is defined as:

\bea \cancel{\hat D}:=\gamma^\mu\hat D_\mu=\gamma^ae^\mu_a\left(\der_\mu-\frac{1}{2}\sigma_{ab}\hat\omega^{\;\;ab}_\mu+\cdots\right).\label{slash-der}\eea In this equation $\gamma^a$ are the (flat) Dirac gamma matrices, $e^a_\mu$ is the tetrad and the ellipsis stands for the missing terms corresponding to the gauge fields $W^{(i)}_\mu$, $B_\mu$ of the gauge group $SU(2)\otimes U(1)$, which are not transformed by the conformal transformation of the metric. In \eqref{slash-der} the Riemannian spin connection $\hat\omega^{\;\;ab}_\mu$ and the commutator of the gamma matrices $\sigma^{ab}$ (the generators of the Lorentz group in the spin representation), read:

\bea &&\hat\omega^{\;\;ab}_\mu=e^{b\nu}\left(\der_\mu e^a_\nu-\{^\lambda_{\mu\nu}\}e^a_\lambda\right),\nonumber\\
&&\sigma^{ab}=\frac{1}{4}\left(\gamma^a\gamma^b-\gamma^b\gamma^a\right),\label{spin-c}\eea respectively. Let us consider the following gauge transformations:

\bea &&g_{\mu\nu}\rightarrow\Omega^2 g_{\mu\nu},\;Q_\mu\rightarrow Q_\mu-2\der_\mu\ln\Omega,\;\psi\rightarrow\Omega^{-3/2}\psi,\nonumber\\
&&e^\mu_a\rightarrow\Omega^{-1}e^\mu_a,\;\gamma^a\rightarrow\gamma^a\Leftrightarrow\sigma^{ab}\rightarrow\sigma^{ab}.\label{gauge-t'}\eea Under the above gauge transformations:

\bea \bar{\psi}\cancel{\hat D}\psi\rightarrow\Omega^{-4}\bar{\psi}\cancel{\hat D}\psi.\nonumber\eea Hence, the Lagrangian $\sqrt{-g}{\cal L}_\text{fermion}$ in its Riemannian form, is already invariant under \eqref{gauge-t'}. In other words, it is not required to make the replacements (the weight $w$ depends on whether the derivative acts on the spinor field or on the tetrad): $$\der_\mu\rightarrow\der^*_\mu=\der_\mu+\frac{w}{2}Q^*_\mu,\;\{^\alpha_{\mu\nu}\}\rightarrow\Gamma^\alpha_{\;\;\mu\nu},$$ in equations \eqref{psi-lag}, \eqref{slash-der} and \eqref{spin-c}, in order for the Lagrangian $\sqrt{-g}{\cal L}_\text{fermion}$ to be gauge invariant. This is true also when the $SU(2)\otimes U(1)$ gauge fields $W^{(i)}_\mu$ and $B_\mu$ are included in the Lagrangian density. This means that the Weyl gauge vector $Q_\alpha$ does not couple neither to fermions nor to other gauge fields including the electromagnetic radiation. This statement has been taken as the basis to avoid the SCE \cite{cheng-1988, cheng-arxiv, tomi-replay}. The argument may be extended to generalized nonmetricity, as shown in \cite{delhom-coupling}. In this reference it has been pointed out that, if take into account an appropriate minimal coupling prescription, the correct Lagrangian density should read:\footnote{In the second term within square brackets the action of the slash operator $\cancel{\hat D}$ should be understood in the following way: $\bar\psi\overleftarrow{\cancel{\hat D}}=(\hat D_\mu\bar\psi)\gamma^\mu$, where the Riemannian gauge derivative $\hat D_\mu$ is defined in \eqref{slash-der}.} 

\bea {\cal L}_\text{fermion}=\frac{i}{2}\left[\bar\psi\left(\cancel{\hat D}\psi\right)-\left(\bar\psi\overleftarrow{\cancel{\hat D}}\right)\psi\right].\label{delhom-lag}\eea 

Let us point out that the above argument is strictly correct only if the mass of the fields $m_\psi$ is assumed vanishing, i. e., if consider the Lagrangian density \eqref{psi-lag}. In this case the exposed argument is just a confirmation of the result discussed at the end of section \ref{sect-geod}, that photons and radiation interact only with the metric field, i. e., with the LC curvature of spacetime. In other words, that these do not interact with nonmetricity. 

Consideration of a point-dependent mass term may radically change the coupling to the nometricity, as anticipated by the form of the geodesic equation \eqref{time-geod'} in $W_4$, where a term $\propto\delta m/\delta x^\alpha$ arises. The hypothesis on the identification of vectors and tensors living in $W_4$ space with physical vectors and tensors, allows to identify any mass parameter with one of the kind \eqref{sce-eq}. As we shall see, this recipe can be applied to Dirac's equation for spinor matter and to the Mathisson-Papapetrou-Dixon equation for the motion of spinning test bodies as well. 

If in place of \eqref{delhom-lag} consider the following Lagrangian density:

\bea {\cal L}_\text{fermion}=\left\{\frac{i}{2}\left[\bar\psi\left(\cancel{\hat D}\psi\right)-\left(\bar\psi\overleftarrow{\cancel{\hat D}}\right)\psi\right]-\bar\psi m_\psi\psi\right\},\label{psi-mass-lag}\eea where $m_\psi\neq 0$, the conclusion of references \cite{cheng-1988, cheng-arxiv} and related works \cite{tomi-replay}, may be incorrect in general. Actually, if assume spaces with arbitrary nonmetricity $W_4$, which means that we must identify the hypothetical vectors and tensors living in $W_4$ with the corresponding physical vectors and tensors, then equation \eqref{sce-eq} must be satisfied. In consequence, the mass of the fermion field $m_\psi$ in \eqref{psi-mass-lag} must obey: 

\bea m_\psi(x)=m_\psi(0)\exp{\left[-\frac{1}{2}\int_{\cal C}Q_{\lambda\mu\nu}u^\mu u^\nu dx^\lambda\right]},\label{m-ferm-w4}\eea which means that there is a non-negligible (under integral) dependence of the mass $m_\psi$ on nonmetricity in \eqref{psi-mass-lag}. 

It seems appropriate to underline, once more, that the identification of the mass of the fermion in \eqref{psi-mass-lag} with the mass in \eqref{m-ferm-w4}, is not an a priori given fact, but it is just a consistency hypothesis or postulate. A similar hypothesis has been made in section \ref{sect-geod} when the factor $\delta m/\delta x^\mu$, appearing in equation \eqref{time-geod'}, was identified with the quantity given by \eqref{dm-eq}. This hypothesis is what makes possible to describe the motion of timelike point particles, including fermion fields, in a gravitational field depicted by the curvature and nonmetricity of $W_4$ space. This is why we call it as ``consistency'' hypothesis. 

This consistency hypothesys may be applied as well to the Mathisson-Papapetrou-Dixon equation, which is the one driving the dynamics of extended, spinning test bodies in curved backgrounds \cite{mathisson, papapetrou, dixon, wald}:

\bea &&\frac{D^*p^\alpha}{ds}=-\frac{1}{2}R^\alpha_{\;\;\mu\nu\lambda}v^\mu S^{\nu\lambda},\nonumber\\
&&\frac{D^*S^{\alpha\beta}}{ds}=2p^{[\alpha}v^{\beta]},\label{mpd-eq}\eea where $p^\alpha$ and $v^\alpha$ are the coordinate components of the four-momentum of the spinning test body and of the unit tangent vector to the worldline $x^\alpha(s)$, respectively.\footnote{Although, in general, four vectors $p^\alpha$ and $mv^\alpha$ differ: $$p^{[\alpha}v^{\beta]}=-\frac{\sqrt{-g}}{4m}\epsilon^{\alpha\beta\mu\nu}R_{\mu\sigma\kappa\lambda}v^\sigma S^{\kappa\lambda}S_\nu,$$ where $S_\alpha$ are the components of the spin vector $$S_\alpha=\frac{\sqrt{-g}}{2m}\epsilon_{\mu\nu\lambda\alpha}p^\mu S^{\nu\lambda},$$ whenever the size of the body is much smaller than the radius of curvature of spacetime, the difference between $p^\alpha$ and $mv^\alpha$ will be negligible compared with $|mv^\alpha|$ itself.} Meanwhile, $S^{\alpha\beta}$ are the components of the spin tensor \cite{wald}. Along the worldline the following condition is satisfied:

\bea g_{\mu\nu}p^\mu S^{\nu\alpha}=0.\label{mpd-cond}\eea where the magnitude $S$ of the spin satisfies: $S^2=S_{\mu\nu}S^{\mu\nu}/2$. Besides, we have that $m^2=-g_{\mu\nu}p^\mu p^\nu$ is the mass (squared) of the spinning body. Hence, if assume the consistency hypothesis, which amounts to identifying the hypothetical four-momentum ${\bf p}$ in \eqref{4-p} with the four-momentum of the spinning test body, its mass will satisfy \eqref{sce-eq}, which is the basis for the SCE.


We have shown that, under the assumption of the parallel transport law \eqref{ptl} and of the consistency hypothesis that allows to identify hypothetical vectors and tensors living in $W_4$, with corresponding physical vectors and tensors, timelike test particles with the mass -- no matter whether these are point particles, spinor fields or spinning bodies -- interact with nonmetricity $Q_{\alpha\mu\nu}$. This means, in turn, that the SCE is inevitable in $W_4$ space. A similar result was obtained in \cite{my-arxiv-paper} under the assumption of a different law of parallel transport.

Our point of view is in clear contradiction with statements found in the bibliography according to which ``the second clock effect has nothing to do with geometry but is entirely determined by the matter coupling'' \cite{hobson, hobson-replay, tomi-replay}. Further investigation of this topic will be the subject of a forthcoming publication \cite{in-prepa}.



\section{Closed timelike worldlines and the Second Clock Effect}\label{sect-ctc}


The demonstration of the absence/occurrence of the SCE in \cite{2clock-tomi} (see appendix \ref{app-a} for the gauge invariant demonstration), as well as in several textbooks \cite{adler-book}, heavily relies on the choice of a closed path which allowed to further apply the Stoke's theorem. The usual argument to justify closed path is that it is necessary in order to check the second clock effect, since the observers have to compare ``notes'' \cite{2clock-tomi}. However, as we have shown above, the SCE arises even if consider paths that are not closed. Observers endowed with identical clocks at the coordinate origin $x=0$, can compare the ticks of their clocks when they coincide again at some distant point $x$, after following different trajectories that joint $0$ and the point $x$. It is not required that the spacetime origin of coordinates and the distant spacetime point coincided as it is for closed paths. As we shall see, given worldline is not closed even if it starting and ending spatial points coincide. For this worldline to be a closed one it is required, besides, that the starting and ending time coordinates coincided as well.

In general, closed paths in spacetime carry causality issues. Timelike worldlines of observers with clocks, aimed at the check of the second clock effect, are not the exception. In this regard we should differentiate the timelike worldlines ${\cal C}$ with coordinates $x^\mu(\xi)$ ($\xi$ is an affine parameter along the worldline), which start and end up at a same spatial point: 

\bea &&x^0(\xi_\text{start})\neq x^0(\xi_\text{end}),\;x^i(\xi_\text{start})=x^i(\xi_\text{end})\nonumber\\
&&\;\;\;\;\;\;\;\;\;\;\;\;\;\;\;\Rightarrow\;x^\mu(\xi_\text{start})\neq x^\mu(\xi_\text{end}),\label{wline-1}\eea from those worldlines $\bar{\cal C}$, which start and end up at the same spacetime point: 

\bea &&x^\mu(\xi_\text{start})=x^\mu(\xi_\text{end})\;\Rightarrow\nonumber\\
&&x^0(\xi_\text{start})=x^0(\xi_\text{end}),\;x^i(\xi_\text{start})=x^i(\xi_\text{end}).\label{wline-2}\eea 

While timelike worldlines of type ${\cal C}$ can be associated with real (classical) motions, timelike worldlines of type $\bar{\cal C}$ are usually called as closed timelike curves (CTCs) and are plagued by causality issues as long as a CTC represents time travel \cite{ctc-morris, ctc-friedman, ctc-thorne, ctc-cho, ctc-bonnor, ctc-luminet}. But integrals of the kind \eqref{12} and \eqref{var-ln'}, which are usually associated with demonstrations about the SCE (see, for instance equation (15.41) of \cite{adler-book} or equation (12) of \cite{2clock-tomi}), go over CTCs as one can easily realize. Hence, these demonstrations are not physically plausible since a physical observer equipped with a clock is a classical system which should respect causality. In general, explanation of the SCE should be based in equations like \eqref{var-ln}, \eqref{master-eq}, \eqref{sce-eq}, or \eqref{app-6} and \eqref{app-6'}, or \eqref{g-ratio} and \eqref{quc}, which involve integration along paths of type $\cal C$, so that, in general terms, Stoke's theorem can not be applied.



\section{Discussion}\label{sect-discu}

Several recent papers have put into discussion the occurrence of the second clock effect \cite{2clock-tomi, tomi-replay, hobson, hobson-replay}. The SCE was the reason why Einstein and others rejected the original work by Weyl in the early twenties of the past century \cite{many-weyl-book}. Resurrection of discussions about Weyl geometry has been fed by the increasing interest in nonmetricity theories and, in particular, in symmetric teleparallel theories of gravity and their application in cosmology. 

The missing ingredient in most papers on nonmetricity theories, including the symmetric teleparallel theories, is gauge symmetry, which is a manifest symmetry of generalized Weyl geometry.\footnote{It is a well-known fact that Weyl geometry spaces are equipped with a conformal structure thanks, precisely, to Weyl gauge symmetry \cite{delhom-2019, many-weyl-book}.} One may wonder whether it makes sense to disregard a manifest symmetry of given geometrical background. The answer may not be unique but, in any case, physical intuition dictates that the existence of a manifest symmetry is not a coincidence that one may overlook without phenomenological consequences. 

Another not very well investigated aspect of nonmetricity theories goes about their geometrical properties and the related phenomenological viability. For instance, the SCE, being a distinctive feature of Weyl geometry spaces, is either just ignored or it is assumed not to occur in physical situations. The basis for neglecting the SCE is the gauge invariance of the Lagrangian for massless Fermions and gauge fields even in its Riemannian version (see, for instance, equation \eqref{psi-lag}), which, strictly speaking, means that the nonmetricity does not interact with massless fields (radiation). 

While the above conclusion is in perfect agreement with our results in this paper: null geodesics in $W_4$ coincide with null geodesics in Riemann space $V_4$, it is not clear why the same conclusion is straightforwardly applied to the case when one adds the mass term as in equation \eqref{psi-mass-lag}. As we have shown, a mass term changes everything. Concluding, for instance, that from Lagrangian density \eqref{psi-mass-lag} and related equations of motion:

\bea \left[i\cancel{\hat D}-m\right]\psi=0,\;\bar\psi\left[i\overleftarrow{\cancel{\hat D}}+m\right]=0,\label{dirac-eq}\eea it follows that fermions and gauge fields do not interact with nonmetricity, amounts to assuming that the mass parameter can not be function (or functional) of the nonmetricity $\bf Q$ with coordinate components $Q_{\alpha\mu\nu}$. However, equation \eqref{sce-eq} is an evidence that for the hypothetical four-momentum living in $W_4$, the mass is a functional which depends on followed path and also on the nonmetricity: $m=m[{\cal C},{\bf Q}]$.

As we have shown in section \ref{sect-fermions}, equations like the geodesics of timelike particles \eqref{time-geod'}, the Dirac equations for spinor fields \eqref{dirac-eq} and the Mathisson-Papapetrou-Dixon equations \eqref{mpd-eq}, which drive the dynamics of spinning test bodies, for non vanishing mass $m\neq 0$, are undetermined until specific postulate or hypothesis on the geometrical nature of the mass is assumed. In the present paper, for instance, the assumed postulate on the nature of the mass parameter is that, the physical quantity and the hypothetical one: the one that appears in the definition of the four-momentum \eqref{4-p} living in $W_4$ space, are to be identified. This identification amounts to equiparate physical vectors and tensors with the related hypothetical vectors and tensors, living in the generalized Weyl space $W_4$. But there are other possibilities.

As an illustration let us assume, as in \cite{hobson, hobson-replay}, that the parallel transport law of given tensor $\bf T$, along the worldline $x^\alpha(\xi)$, reads:

\bea \frac{D^*{\bf T}}{d\xi}=\frac{dx^\mu}{d\xi}\nabla^*_\mu{\bf T}=0,\label{ill-1}\eea which means that Weyl gauge symmetry is being considered as a manifest symmetry of generalized Weyl spaces $W_4$, and that the following hypothesis on the nature of mass holds:

\bea m=m_0\vphi,\label{ill-2}\eea where $m_0$ is a constant and $\vphi$ is a scalar field. 

Let us first assume that the mass \eqref{ill-2} is the one that appears in the definition of the hypothetical four-momentum \eqref{4-p}. Then, since $-m^2=g_{\mu\nu}p^\mu p^\nu$, the following chain of equations takes place:

\bea &&-\frac{D^*m^2}{d\xi}=\frac{dx^\lambda}{d\xi}\nabla^*_\lambda g_{\mu\nu}p^\mu p^\nu\nonumber\\
&&\;\;\;\;\;\;\Rightarrow\;\frac{d\ln m}{d\xi}=\frac{dx^\lambda}{d\xi}\left(Q^*_\lambda+\frac{1}{2}Q_{\lambda\mu\nu}u^\mu u^\nu\right)\nonumber\\
&&\;\;\;\;\;\;\Rightarrow\;\der_\alpha\ln\vphi-\left(Q^*_\alpha+\frac{1}{2}Q_{\alpha\mu\nu}u^\mu u^\nu\right)=0.\nonumber\eea From the last equation it follows that:

\bea Q_{\alpha\mu\nu}=2\left(Q^*_\alpha-\der_\alpha\ln\vphi\right)g_{\mu\nu}.\nonumber\eea This case corresponds to vectorial nonmetricity with gauge vector $Q_\alpha=2\left(Q^*_\alpha-\der_\alpha\ln\vphi\right)$, i. e., it corresponds to standard Weyl space $\tilde W_4$. Hence, the assumption of the parallel transport law \eqref{ill-1} and that the mass $m$ in \eqref{ill-2} is to be identified with the mass in the definition of the hypothetical four-momentum \eqref{4-p}, singles out $\tilde W_4$ space as the geometric arena where the gravitational laws take place. This conclusion is independent of the specific gravitational theory considered. 

Let us now assume that \eqref{ill-1} and \eqref{ill-2} take place, but the mass $m$ is not the one that appears in the definition of the four-momentum $\bf p$ in \eqref{4-p}. This means that we are renouncing to associate physical vectors (and tensors) with hypothetical vectors (and tensors) that live in $W_4$ (or in any of its subclasses). In consequence, we are renouncing to describe the given gravitational laws in generalized Weyl space with arbitrary nonmetricity.

Consideration of given gravitational action $S_g=\int d^4x\sqrt{-g}{\cal L}_g$, where ${\cal L}_g$ is the gravitational Lagrangian density, adds additional possibilities. One may consider, for instance, a gauge invariant gravitational Lagrangian $\sqrt{-g}{\cal L}_g$, so that the derived gravitational equations will respect the manifest symmetry of background space $W_4$. Or one may, alternatively, consider a gravitational Lagrangian without gauge symmetry, even if the geometric background space $W_4$ is gauge symmetric. This, of course, will lead to a gravitational theory that is not gauge invariant. In this last case the manifest symmetry of the geometric background (gauge symmetry) is underutilized and may be ignored. In consequence, gauge invariant derivative operators may be replaced by non-gauge invariant ones: $\nabla^*_\alpha\rightarrow\nabla_\alpha$, $D^*/d\xi\rightarrow D/d\xi$, etc.

The lesson to be learned is that definitive conclusions on a given setup can not be given until the underlying postulates and/or assumptions are clearly stated. Conclusions such as either: ``The SCE does not take place...'' or ``The SCE is inevitable...'', require to be complemented with statements about the underlying assumptions: ``...if take into account the following assumptions...''



\section{Conclusion}\label{sect-conclu}

In this paper we have demonstrated that, if assume that: (i) gauge symmetry is a manifest symmetry of generalized Weyl space, (ii) the parallel transport law \eqref{ptl} holds in $W_4$ and (iii) identification of physical vectors and tensors with the related hypothetical vectors and tensors living in $W_4$ takes place, the second clock effect is inevitable. In consequence, under the above assumptions, gauge invariant theories that are based in $W_4$ background spaces, are phenomenologically ruled out. Only in the subclass of $W_4$ known as Weyl integrable geometry, the SCE does not take place. Hence WIG is the only phenomenologically viable non-Riemannian gauge invariant background space from the classical perspective. 


Although generalizations of nonmetricity recently investigated within the framework of gauge invariant teleparallel theories of gravity, are phenomenologically ruled out in the classical context, in the domain of quantum gravity these may play a fundamental role. This and related issues are the subject of our current work.


{\bf Acknowledgments.} The author thanks M P Hobson, A N Lasenby and M Adak for their useful comments, and also A Delhom and M Adak for pointing out the references \cite{delhom-2019} and \cite{tucker}, respectively. I acknowledge FORDECYT-PRONACES-CONACYT for support of the present research under grant CF-MG-2558591.


\appendix




\section{Gauge symmetry, closed paths and second clock effect}\label{app-a}

In this appendix section we shall discuss on what happens if consider gauge symmetry in a situation like the one discussed in lemma 2 of \cite{2clock-tomi}. For simplicity, instead of parallel transport of the inner product of two vectors we shall consider parallel transport of the length of a given vector.

Consider parallel transport of a vector ${\bf v}$, with coordinate components $v^\alpha$ and conformal weight $w({\bf v})=w$, along a closed path $\bar{\cal C}$. In this case equation \eqref{var-ln} is rewritten in the following form: 

\bea \Delta\ln v(x)=-\frac{w+1}{2}\oint_{\bar{\cal C}}Q_{\lambda\mu\nu}t^\mu t^\nu dx^\lambda,\label{var-ln'}\eea where $v\equiv||{\bf v}||$ is the length of vector ${\bf v}$. Following \cite{2clock-tomi}, we apply the Stoke's theorem to the path integral \eqref{var-ln'}, we get:

\bea &&\oint_{\bar{\cal C}}Q_{\lambda\mu\nu}t^\mu t^\nu dx^\lambda=\iint_S\nabla_{[\lambda}\left(Q_{\sigma]\mu\nu}t^\mu t^\nu\right)dx^\lambda\wedge dx^\sigma\nonumber\\
&&=\iint_S\left\{\nabla_{[\lambda}Q_{\sigma]\mu\nu}t^\mu t^\nu+\nabla_{[\lambda}\left(t^\mu t^\nu\right)Q_{\sigma]\mu\nu}\right\}dx^\lambda\wedge dx^\sigma\nonumber\\
&&=\iint_S\left\{R_{(\mu\nu)\lambda\sigma}t^\mu t^\nu+\nabla_{[\lambda}\left(t^\mu t^\nu\right)Q_{\sigma]\mu\nu}\right\}dx^\lambda\wedge dx^\sigma,\nonumber\eea where $S$ is any surface with boundary $\bar{\cal C}$ and we have taken into account the third Bianchi identity \eqref{3-bianchi-id}. Consider further the dagger derivative \eqref{dag-der-v}, \eqref{dag-der-t}:

\bea \nabla^\dag_\lambda\left(t^\mu t^\nu\right)=\nabla_\lambda\left(t^\mu t^\nu\right)-\frac{1}{2}Q^{\;\;\mu}_{\lambda\;\;\kappa}t^\kappa t^\nu-\frac{1}{2}Q^{\;\;\nu}_{\lambda\;\;\kappa} t^\kappa t^\mu,\nonumber\eea and take into account that the unit vector ${\bf t}$ is parallel transported along the closed path $\bar{\cal C}$, then:

\bea \nabla^\dag_\lambda\left(t^\mu t^\nu\right)=0\Rightarrow\nabla_\lambda\left(t^\mu t^\nu\right)=Q^{\;\;(\mu}_{\lambda\;\;\;\kappa}t^{\nu)} t^\kappa.\label{app-7}\eea Hence, the above path integral along the closed worldline $\bar{\cal C}$ can be written as it follows:

\bea &&\oint_{\bar{\cal C}}Q_{\lambda\mu\nu}t^\mu t^\nu dx^\lambda=\nonumber\\
&&\iint_S\left\{R_{(\mu\nu)\lambda\sigma}+Q^{\;\;\;\kappa}_{\lambda\;\;(\mu}Q_{\sigma\kappa\nu)}\right\}t^\mu t^\nu dx^\lambda\wedge dx^\sigma.\label{app-9}\eea Equation \eqref{var-ln'} can then be written in the following way (compare with \eqref{15-correct}):

\bea &&\Delta\ln v(x)=-\frac{w+1}{2}\iint_S\left\{R_{(\mu\nu)\lambda\sigma}\right.\nonumber\\
&&\left.\;\;\;\;\;\;\;\;\;\;\;\;\;\;\;\;\;\;\;\;\;+Q^{\;\;\;\kappa}_{\lambda\;\;(\mu}Q_{\sigma\kappa\nu)}\right\}t^\mu t^\nu dx^\lambda\wedge dx^\sigma,\label{app-fin}\eea where, we recall, $w$ is the conformal weight of vector ${\bf v}$.

We have demonstrated that the length of vectors is path dependent even if consider the teleparallel condition $R_{\mu\nu\lambda\sigma}=0$. This result is contrary to the statement of lemma 2 of \cite{2clock-tomi}. Notice that if ignore the teleparallel condition in $\tilde W_4$, since $Q_{\alpha\mu\nu}=Q_\alpha g_{\mu\nu}$, equation \eqref{app-fin} transforms into the well-known equation:

\bea \Delta\ln v(x)=-\frac{w+1}{2}\iint_S\der_{[\lambda}Q_{\sigma]}dx^\lambda\wedge dx^\sigma.\label{app-fin'}\eea



\section{Short reply to a comment in reference \cite{tomi-replay}}\label{app-b}

In section IV.A of reference \cite{tomi-replay} it is stated that: 

\begin{itemize}

\item ``...then modifies the theory in order to implement the ``manifest symmetry'', according to which spacetime vectors $V^\mu$ transform with some weight $\alpha$ as $V^\mu\rightarrow e^{\alpha\phi}V^\mu$. The basic error in this argument is that the conformal transformation is explicit for the tangent space vectors $V^a\rightarrow e^\phi V^a$, and therefore spacetime vectors $V^\mu=e^\mu_{\;\;a}V^a$ should consistently have the zero weight $\alpha=0$.''

\end{itemize} The comment refers to our paper \cite{my-arxiv-paper} and also to a previous version of the present work. Before replying to this comment, in order to unify notations, here we make the following replacements: the conformal factor $\Omega\rightarrow e^\phi$, while the conformal weight of given tensor $w\rightarrow\alpha$. 

According to the above comment it is a basic error to consider vectors $V^\mu$ with conformal weight $\alpha\neq 0$. Hence, our straightforward reply is to show that, indeed, there are vectors with conformal weight $\alpha\neq 0$. Take, for instance, the timelike four-velocity vector: $u^\mu=dx^\mu/ds$ ($ds$ is the arc-length), which is tangent to the worldline $x^\mu(s)$. Under a conformal transformation of the kind we consider in our papers \cite{my-arxiv-paper} and in the present work (see footnote \ref{foot-1} in the main text of this paper): 
 
\bea g_{\mu\nu}\rightarrow e^{2\phi}g_{\mu\nu},\;dx^\mu\rightarrow dx^\mu\;\Rightarrow\;ds\rightarrow e^\phi ds,\label{conf-transf}\eea the components of the four-velocity transform like:

\bea u^\mu\rightarrow e^{-\phi}u^\mu,\label{4-vel-t}\eea so that its weight $\alpha(u^\mu)=-1$. In consequence, the timelike four-momentum $p^\mu=m u^\mu$, where the mass $m$ of given point-particle has weight $\alpha(m)=-1$, has conformal weight $\alpha(p^\mu)=-2$. 

In general, given any vector $V^\mu$ with length 

\bea V=\sqrt{g_{\mu\nu}V^\mu V^\nu},\label{vect-l}\eea and conformal weight $\alpha$, one can define the spacelike unit vector with coordinate components:

\bea t^\mu=\frac{V^\mu}{V}\;\Rightarrow\;g_{\mu\nu}t^\mu t^\nu=1.\label{slk-u-vect}\eea Since $\alpha(V^\mu)=\alpha$, while $\alpha(g_{\mu\nu})=2$, then $\alpha(V)=\alpha+1$. Means that $\alpha(t^\mu)=-1$ independent of the conformal weight of vector $V^\mu$. As an illustration let us assume, as stated in the mentioned comment in \cite{tomi-replay}, that $\alpha(V^\mu)=\alpha=0$. Then, from \eqref{vect-l} it follows that $\alpha(V)=1$, so that, taking into account \eqref{slk-u-vect}, one gets that, $\alpha(t^\mu)=-1$. This example shows that, no matter whether we have vector fields with vanishing conformal weight, one can always construct spacelike unit vectors with weight $\alpha=-1$. Additionally, one can find in the bibliography quite the contrary statement regarding tangent space vectors $V^a$, whose conformal weight is taken to be vanishing \cite{hobson, hobson-replay}: $\alpha(V^a)=0$. 

The above examples: the timelike four-velocity and four-momentum vectors, as well as the spacelike unit vector \eqref{slk-u-vect}, show that the comment in section IV.A of reference \cite{tomi-replay} is either simply incorrect, or the authors of that reference are considering conformal transformations of a different kind to the one considered in \cite{my-arxiv-paper} and in the present paper.




\end{document}